\documentclass[10pt]{iopart}

\makeatletter
\expandafter\let\csname equation*\endcsname\@undefined
\expandafter\let\csname endequation*\endcsname\@undefined
\makeatother

\usepackage{amsmath, amssymb, amsfonts, amsthm}
\usepackage{mathtools}
\usepackage{bbm}
\usepackage{dsfont}
\usepackage{latexsym}
\usepackage{xcolor}

\usepackage{physics}
\newcommand{\im}{\mathbbm{i}}

\usepackage{graphicx}
\usepackage{epstopdf}
\AppendGraphicsExtensions{.tif}

\usepackage{booktabs}
\usepackage{multirow}
\usepackage{tabularx}

\usepackage{caption}
\usepackage{subcaption}

\theoremstyle{definition}

\newtheorem{remark}{Remark}[section]

\usepackage{nicefrac}
\usepackage[english]{babel}
\usepackage[T1]{fontenc}
\usepackage[utf8]{inputenc}
\usepackage{footmisc}
\usepackage{iopams}

\begin{document}

\title[]{de Broglie-Bohm Dynamics with Schrödinger Source Fields: A Framework for Subquantum Theory}

\author{Said Mikki}

\address{Zhejiang University/University of Illinois at Urbana-Champaign (ZJU-UIUC) Institute, Zhejiang University, Haining, Zhejiang, China}
\ead{smikki@illinois.com}

\vspace{10pt}

\begin{abstract}
Subquantum pilot-wave theory treats the Born rule not as a postulate but as a statistical equilibrium, toward which nonequilibrium distributions relax under coarse-graining. The relaxation is passive, and its fine-grained structure is frozen: the ratio between the particle density and the Born density is conserved along every Bohmian trajectory. We show that this constraint is an artefact of the source-free Schr\"odinger equation. Introducing a complex source field yields a consistent extended subquantum dynamics in which the source couples directly to the pilot wave and the ratio field obeys an exact transport equation with a source term, replacing the conservation law at the fine-grained level. The resulting subquantum theory formally contains standard quantum mechanics as the equilibrium sector of vanishing source and Born-distributed particles. Quantum equilibrium thereby ceases to be an inescapable attractor and becomes a controllable regime of a broader subquantum dynamics.
\end{abstract}

\section{Introduction}
\label{sec:introduction}

The concept of the field, introduced by Faraday in the first half of the nineteenth century, underwent its second great transformation in the first quarter of the twentieth, when the idea of matter waves extended the domain of field theory from electromagnetism to the description of material particles. In one of the most consequential moments in the history of physics, Louis de Broglie proposed in his 1924 doctoral dissertation a new class of fields---the ``matter waves'' represented by functions $\psi(\vb{x})$ on position space $\vb{x} \in \mathbb{R}^3$ \cite{Bacciagaluppi_Valantini2009,deBroglie1924}. Within two years, Erwin Schr\"odinger took up de Broglie's idea and developed it into wave mechanics, an alternative to the matrix mechanics of Born, Jordan, and Heisenberg \cite{Schrodinger2020collected_papers}. In the standard historical narrative, de Broglie's contributions were largely subsumed under Schr\"odinger's revolutionary achievements \cite{Duncan2023volume_2_1923-1927,jammer1974the_philosophy_of_quantum_mechanics}.
Yet Schr\"odinger's fields are by no means identical to de Broglie's \cite{Valentini2025beyond_the_quantum}. As recent scholarship has made increasingly clear, de Broglie's original conception was broader in scope and deeper in its ontological commitments. He sought to reformulate the foundations of dynamics itself. His fundamental fields were not conceived merely as analogues of the electromagnetic field, but as a universal feature of all matter-energy processes \cite{Bacciagaluppi_Valantini2009}. The Schr\"odinger equation, in de Broglie's vision, was to govern the dynamics of this overarching field-theoretic framework. Schr\"odinger's formulation, however, eliminated an essential component of de Broglie's theory: the particle. For de Broglie, particles were fundamental. His theory is irreducibly dualistic, positing that both fields and particles are indispensable to any complete description of physical reality. The de Broglie field $\psi$ functions as a guiding wave that pilots the particle; the particle itself is real and belongs to the primitive ontology of the world. Thus, pilot-wave theory postulates at least two fundamental dynamical fields: the pilot-wave field $\psi(\vb{x},t)$, which guides the particles, and the probability distribution $P(\vb{x},t)$, which encodes the statistical density of the particle ensemble. The former embodies the field-theoretic aspect of the dualistic ontology \cite{oldofredi2023orthodox}; the latter captures its particle-like dimension.\footnote{Although the standard dBB dynamical theory posits particles \emph{and} fields on an equal ontological footing, its mathematical formulation is predominantly field-theoretic, with $P$ arising from the ensemble treatment of statistical mechanics \cite{tolman1979the_Principles_of_Statistical_Mechanics}. Moreover, de Broglie himself later turned to developing a monist field-like theory, the ``double-solution'' theory \cite{deBroglie1960nonlinear,BohmHiley1982}, which attempts to replace a particle-like field such as the probability distribution $P$ with a pure $u$-field whose sole purpose is to generate particles through dynamical singularities \cite{oldofredi2025pilotwave}. De Broglie later abandoned this approach, but it has been revived several times by other researchers; see, e.g., \cite{Holland2019,Drezet2020guidance}. The extended dBB formalism proposed in the present paper is motivated by the dualistic version of dBB theory, not the monist version.}

This dualistic ontology was rediscovered by David Bohm in 1952, who formulated a distinct pilot-wave theory by introducing the quantum potential and exploiting the Hamilton-Jacobi structure of the Schr\"odinger equation \cite{Bohm1952Part_I,Bohm1952Part_II}. Although the de Broglie and Bohm formulations can be made to yield identical empirical predictions, they are not equivalent as physical theories \cite{Colin_valentini2014}. For Schr\"odinger, the field alone was ontologically primary; his elimination of particles gave rise, inevitably, to the measurement problem. In the de Broglie-Bohm (dBB) theory, by contrast, the measurement problem dissolves: measurement outcomes are directly correlated with the actual trajectories of particles---for instance, the positions of pointer states \cite{Bohm1952Part_I,Bohm1952Part_II,bohm1993the_undivided_universe}.

In a series of papers beginning in the early 1990s, Valentini revived the concept of quantum nonequilibrium by demonstrating that the Born rule $P = |\psi|^2$ need not be a fundamental postulate. It can instead be understood as a state of statistical equilibrium, analogous to thermal equilibrium in classical statistical mechanics \cite{Valentini1991a,Valentini1991b}. Central to this formulation is a subquantum entropy and an associated $H$-theorem, which shows that an arbitrary initial distribution $P$ relaxes toward the Born distribution under coarse-graining, much as a classical gas relaxes toward Maxwell-Boltzmann equilibrium. This framework has significant implications: quantum mechanics as we know it may be only the equilibrium limit of a deeper, nonequilibrium theory; deviations from the Born rule---if they could be prepared or sustained---would lead to violations of the uncertainty principle, superluminal signalling, and other phenomena beyond the reach of standard quantum mechanics \cite{Valentini1991a,Valentini1991b}. Subsequent numerical and analytical investigations have extensively explored the timescales and mechanisms of this passive relaxation to quantum equilibrium in source-free dBB theory \cite{ColinStruyve2010,TowlerRussellValentini2012,LustosaColinBergliaffa2021,LustosaPintoNetoValentini2023}.

Our primary aim in this paper is to reexamine dBB dynamics from a perspective prompted by a question that initially appears marginal: Why does the de Broglie field $\psi$ in the Schr\"odinger equation lack a source term? The asymmetry between quantum mechanics and classical field theory in terms of the status of a source field is striking. In the paradigmatic classical field theories---acoustics, electromagnetism, Newtonian gravity, general relativity---the dynamical field of interest is generated by an associated source.\footnote{The claim is not universal: nonlinear and topological field theories such as the Sine--Gordon model, pure Yang--Mills theory, and vacuum general relativity do not require an externally prescribed source; their nontrivial field configurations arise instead from topology, self-coupling, or nonlinearity. The contrast with quantum mechanics is drawn here with respect to the paradigmatic sourced field theories that historically motivated the field concept.}
The absence of a source term in the Schr\"odinger equation is normally regarded as mandatory because $|\psi|^2$ is interpreted as the probability density: any source would violate probability conservation. In the de Broglie-Bohm ontology, however, this argument does not apply at the fundamental level, since the physical probability distribution is $P$, while $\psi$ is an ontological guiding field. Probability is vested in the particle distribution $P$, whose continuity equation remains source-free irrespective of $J$; the Born density $|\psi|^2$ need not be conserved because it is not a probability density. The objection therefore rests on a conflation that the dBB ontology rejects, and a consistent dynamical theory of sourced de Broglie fields becomes possible.

We argue that if de Broglie's program of a universal pilot-wave field theory is to succeed, sources should appear naturally within the formalism. We introduce a complex Schr\"odinger source field $J(\vb{x},t)$ and develop the resulting \emph{extended dBB dynamics}. The framework operates with three fundamental fields---the pilot-wave $\psi$, the particle distribution $P$, and the source $J$. In the conventional source-free framework, $P$ is transported along the Bohmian trajectories defined by $\psi$, and the ratio $f = P/|\psi|^2$ is strictly conserved along these trajectories. Valentini exploited this conservative dynamics to develop a subquantum $H$-theorem, but only at the coarse-grained level, since at the fine-grained level $f$ cannot change \cite{Valentini1991a,Valentini2025The_trouble}. The introduction of $J$ changes this picture. The source modifies the pilot-wave (reshaping both its amplitude and phase), which in turn alters the velocity field that transports $P$; moreover, as we show, the evolution equation for $f$ acquires a source/sink term driven by $J$, so that $f$ is no longer constant along trajectories. This term can locally increase or decrease $f$ in a controlled manner.

Because $J$ breaks the Lagrangian invariance of $f$, the fine-grained entropy drifts, and relaxation toward or away from equilibrium can be described without invoking coarse-graining. For the source-driven part of the dynamics this sidesteps the re-randomization issues of coarse-graining derivations \cite{Callender2007}, and since the source breaks the equivariance of the Born density, typicality arguments \cite{durrr2009bohmian_mechanics} do not apply to it unchanged.

The source-free dynamics thus reduces to a closed, conservative system whose only nonequilibrium behaviour is a coarse-grained relaxation toward equilibrium. At the fine-grained level an initial mismatch is merely carried along the flow, since $f$ is conserved, and the only mechanism that can change the statistics is coarse-grained relaxation. The source-free theory therefore cannot actively drive the system toward or away from equilibrium, nor reshape the nonequilibrium profile, as an open dynamical system can. It is precisely this limitation that motivates the introduction of $J$.

The scope of the paper merits an explicit statement. The extended dBB theory is a \emph{subquantum} theory: it subsumes standard quantum mechanics as the special case $J=0$ with $P=|\psi|^2$ at the initial time, and it is concerned with departures from Born-like statistics and with the dynamical origin of equilibrium. Its intent therefore differs fundamentally from that of collapse models such as GRW and CSL, in which the Schr\"odinger equation is modified expressly to reproduce the Born rule. Here the Born rule is not the object of the modification but one regime of a broader dynamics (Sec.~\ref{sec:why_born_rule} and \ref{app:source_comparisons}).

A second, related motivation arises from the status of quantum equilibrium. The dBB community often treats it as a global attractor, with all nonequilibrium distributions eventually converging to the Born rule (``quantum death'' \cite{Valentini2025beyond_the_quantum}). This convergence, however, is established at the coarse-grained level: in the source-free theory the subquantum $H$-theorem guarantees relaxation under suitable statistical mixing conditions, whereas at the fine-grained level the ratio $f$ is a strict Lagrangian invariant, so no relaxation occurs at that level \cite{Valentini1991a}. The extended framework shows that the Born distribution need not be an inescapable attractor \emph{even at the fine-grained level}: by suitable choice of $J$, the system can be driven into, out of, or sustained away from quantum equilibrium. Because $J$ is complex, its two projections onto the instantaneous phase of $\psi$ govern different aspects of the dynamics---the phase (trajectories and effective quantum potential) and the amplitude (the ratio $f$ and the fine-grained entropy)---enabling independent control over both.

The introduction of $J$ as an external source term, rather than as a nonlinear modification of the Schr\"odinger equation, is a deliberate methodological choice with substantive consequences. A nonlinear correction designed to imitate source-like behaviour can usually be absorbed into an effective internal source, blurring the distinction between a dynamical field and its generator. An externally prescribed $J$ leaves the equation affine in $\psi_J$ and preserves that distinction: the general solution remains the unitary evolution of the initial state plus a particular integral over $J$, and the source may be specified independently of the state on which it acts (Duhamel principle \cite{cazenave2003semilinear}). A single formalism thereby accommodates both a phenomenological reading, in which $J$ encodes external control fields or unresolved environmental degrees of freedom, and an ontological reading, in which $J$ is a fundamental subquantum field. The present paper develops the dynamics common to both and does not adjudicate between them.

\section{de Broglie-Bohm (dBB) Dynamics}
\label{sec: de Broglie-Bohm Dynamics}

We begin by reviewing the essential elements of dBB pilot-wave theory.

\subsection{Madelung Transformation and the Guidance Equation}
\label{sec:madelung_transformation}

The wavefunction (the de Broglie pilot-wave field $\psi$) is recast via the Madelung transformation \cite{Madelung1927} into its hydrodynamic form by writing it in polar form as
\begin{equation}
\psi(\vb{x},t) = R(\vb{x},t) e^{\im S(\vb{x},t)/\hbar},
\label{eq:polar}
\end{equation}
where $R(\vb{x},t)$ and $S(\vb{x},t)$ are real-valued scalar fields. Substituting this into the standard Schr\"odinger equation yields two coupled equations: the quantum Hamilton-Jacobi equation
\begin{equation}
\frac{\partial S(\vb{x},t)}{\partial t} + \frac{[\nabla S(\vb{x},t)]^2}{2m} + V(\vb{x},t) - \frac{\hbar^2}{2m} \frac{\nabla^2 R(\vb{x},t)}{R(\vb{x},t)} = 0,
\label{eq:hamilton_jacobi}
\end{equation}
and the continuity equation
\begin{equation}
\frac{\partial |\psi(\vb{x},t)|^2}{\partial t} + \nabla \cdot \left( |\psi(\vb{x},t)|^2 \frac{\nabla S(\vb{x},t)}{m} \right) = 0.
\label{eq:continuity_born}
\end{equation}
The second equation, Eq.~\eqref{eq:continuity_born}, is readily identified as a continuity equation (conservation law) for the quantity $|\psi(\vb{x},t)|^2$, with associated flux
\begin{equation}\label{eq: quantum probablity current}
\vb{j}(\vb{x},t) = \frac{|\psi(\vb{x},t)|^2}{m} \nabla S(\vb{x},t).
\end{equation}
This observation naturally suggests that if underlying particles exist, they should move with a velocity that generates the current $\vb{j}(\vb{x},t)$, namely $\vb{v}(\vb{x},t) = \nabla S(\vb{x},t)/m$, where $m$ is the particle mass. Thus, a velocity field is associated with the particle motion, introducing the concept of a flow.\footnote{Madelung had earlier introduced the first hydrodynamic formulation of quantum mechanics, in which the Schr\"odinger equation was transformed into a set of fluid-dynamical equations \cite{Madelung1927}.}

Accordingly, de Broglie advanced the following proposal as the fundamental law of motion:
\begin{equation}
\frac{\dd \vb{q}(t)}{\dd t} = \vb{v}^{\psi}[\vb{q}(t), t] := \frac{\vb{j}[\vb{q}(t), t]}{|\psi[\vb{q}(t), t]|^2} = \frac{1}{m} \nabla S[\vb{q}(t), t],
\label{eq:guidance}
\end{equation}
where $\vb{j}(\vb{x},t)$ is the quantum probability current defined in \eqref{eq: quantum probablity current}. Here we have introduced the notation $\vb{q}(t)$ to denote the trajectory of a specific Bohmian particle, thereby distinguishing it from the field variable $\vb{x}$, which serves as the independent coordinate in the configuration space. In this formulation, de Broglie dynamics constitutes a first-order theory of motion.

Before presenting the key ideas of quantum nonequilibrium, we recall that the derivation of the continuity equation for $P(\vb{x},t)$ rests on two assumptions intrinsic to the dBB ontology \cite{Bohm1953,Valentini1991a,Valentini2025The_trouble}:
\begin{enumerate}
\item \emph{Particle Conservation:} Particles are conserved entities; they do not appear or disappear from the configuration space.
\item \emph{Deterministic Velocity Field:} Each particle moves with a well-defined, single-valued velocity field $\vb{v}^{\psi}(\vb{x},t)$ given by the guidance equation \eqref{eq:guidance}, which is smooth enough to guarantee existence and uniqueness of trajectories.
\end{enumerate}
The continuity equation for $P(\vb{x},t)$ then follows:
\begin{equation}
\frac{\partial P(\vb{x},t)}{\partial t} + \nabla \cdot \bigl[ P(\vb{x},t) \, \vb{v}^{\psi}(\vb{x},t) \bigr] = 0.
\label{eq:continuity_P}
\end{equation}
This equation is purely kinematic and does not require the Born rule $P = |\psi|^2$. The Born density $|\psi|^2$ satisfies an analogous equation with the same velocity field \cite{Valentini1991a}:
\begin{equation}
\frac{\partial |\psi(\vb{x},t)|^2}{\partial t} + \nabla \cdot \bigl[ |\psi(\vb{x},t)|^2 \, \vb{v}^{\psi}(\vb{x},t) \bigr] = 0.
\label{eq:continuity_born_2}
\end{equation}
Defining the ratio \cite{Bohm1953}
\begin{equation}
f(\vb{x},t) := \frac{P(\vb{x},t)}{|\psi(\vb{x},t)|^2},
\end{equation}
and subtracting \eqref{eq:continuity_born_2} from \eqref{eq:continuity_P}, we find that $f$ is advected along the flow \cite{Rajeev2018FluidMechanics}:
\begin{equation}
\frac{D f(\vb{x},t)}{D t} := \frac{\partial f(\vb{x},t)}{\partial t} + \vb{v}^{\psi}(\vb{x},t) \cdot \nabla f(\vb{x},t) = 0.
\label{eq:f_advected}
\end{equation}
Equation \eqref{eq:f_advected} implies that $f$ is constant along each Bohmian trajectory $\vb{q}(t)$. Consequently, if $P(\vb{x},0) = |\psi(\vb{x},0)|^2$ holds initially, it holds for all time; conversely, a deviation present initially persists along trajectories in the absence of additional relaxation mechanisms. As noted by Bohm \cite{Bohm1953}, \eqref{eq:f_advected} is analogous to the Liouville equation in Hamiltonian statistical mechanics, and it implies that it suffices for the initial distribution to satisfy $P(\vb{x},0) = |\psi(\vb{x},0)|^2$ for dBB dynamics to reproduce the predictions of standard quantum mechanics \cite{Bohm1952Part_I,Bohm1952Part_II,bohm1993the_undivided_universe}. As will be shown below, when a source term is admitted into the Schr\"odinger equation, this result no longer holds.

By contrast, Valentini formulated a subquantum $H$-theorem showing that, under coarse-graining and for a wide class of initial states, a source-free dBB system can relax toward the Born distribution \cite{Valentini1991a, ValentiniWestman2005}. Numerical studies, including interacting systems such as coupled harmonic oscillators \cite{TowlerRussellValentini2012, ColinStruyve2010, LustosaColinBergliaffa2021, LustosaPintoNetoValentini2023}, show that this relaxation is sensitive to the system's dynamics: interactions or a low number of superposed modes can significantly delay it and leave persistent nonequilibrium residues \cite{AbrahamColinValentini2014}. The source-free theory thus already admits persistent nonequilibrium, which motivates the exploration of broader dynamical regimes.

\section{Modified Schr\"odinger Equation with Source Terms}
\label{sec: Modified Schrodinger Equation with Source Terms}

We now seek to expand the original program of subquantum theory by introducing a complex source field $J(\vb{x},t)$ that excites the Schr\"odinger equation. This extension turns subquantum dynamics from a closed, conservative system into an open one. In the following we derive the modified Hamilton-Jacobi and continuity equations.

\subsection{Modified Hamilton-Jacobi and Continuity Equations}
\label{sec:modified_HJ_continuity}

We consider the modified Schr\"odinger equation with a complex source term, which we dub the Schr\"odinger source field. Formally, this is a time-dependent complex-valued scalar field defined on the configuration space $\mathcal{C}$ of the system under consideration:
$J: \mathcal{C} \times \mathbb{R} \times \mathcal{O} \to \mathbb{C},$
parameterized by the Hamiltonian $\hat{H} \in \mathcal{O}$, where $\mathcal{O}$ is the set of well-defined Hamiltonian operators of the theory. In plain terms, \(J\) is an exogenous complex-valued scalar field: in the
forward initial-value problem it is prescribed independently of the
wavefunction \(\psi\), may depend on the Hamiltonian, and is not required to
be a functional of \(\psi\). We assume that $J(\vb{x},t;\hat{H})$ possesses sufficient regularity to ensure that the modified Schr\"odinger equation admits a unique, well-behaved solution. For concreteness, we consider a particle of mass $m$ in a potential $V(\vb{x},t)$:
\begin{equation}
\im \hbar \frac{\partial \psi_J(\vb{x},t)}{\partial t} = -\frac{\hbar^2}{2m} \nabla^2 \psi_J(\vb{x},t) + V(\vb{x},t) \psi_J(\vb{x},t) + J(\vb{x},t),
\label{eq:modified_schrodinger}
\end{equation}
The solution implies that the de Broglie field is a functional of the source, $\psi = \mathcal{F}[J]$, or $\psi_J(\vb{x},t)$ when the spacetime dependence is made explicit. We adopt the subscript notation throughout to highlight this dependence. All field quantities inherit it, including the velocity field $\vb{v}^{\psi_J}(\vb{x},t) := \nabla S_J(\vb{x},t)/m$ and the Born density $P^{\psi_J}_{\rm B}(\vb{x},t) := |\psi_J(\vb{x},t)|^2$. Conventional quantum mechanics is recovered in the limit $J = 0$.\footnote{For background on source-based modifications of the Schr\"odinger equation, see \ref{app:source_comparisons}. The constructions discussed there---the double-solution program \cite{Holland2019}, particle-centered action-reaction models \cite{Sutherland2019a,Sutherland2019b,Darrow2024}, non-Hermitian quantum dynamics \cite{Bender2007,Moiseyev2011}, and collapse models---pursue different foundational objectives and are conceptually distinct from the present framework.}

Writing $\psi_J(\vb{x},t) = R_J(\vb{x},t) e^{\im S_J(\vb{x},t)/\hbar}$ and decomposing the source into real and imaginary parts,
\begin{equation}
J(\vb{x},t) = J_{\mathrm{r}}(\vb{x},t) + \im J_{\mathrm{i}}(\vb{x},t),
\label{eq:source_decomp}
\end{equation}
we obtain the modified Hamilton-Jacobi equation
\begin{equation}
\boxed{
\begin{aligned}
\frac{\partial S_J(\vb{x},t)}{\partial t} + \frac{\bigl[\nabla S_J(\vb{x},t)\bigr]^2}{2m} + V(\vb{x},t) + Q^{\rm B}_J(\vb{x},t)
&= -\frac{1}{R_J(\vb{x},t)} \operatorname{Re}\left[ J(\vb{x},t) e^{-\im S_J(\vb{x},t)/\hbar} \right] ,
\end{aligned}
}
\label{eq:modified_HJ}
\end{equation}
where
\begin{equation}
Q^{\rm B}_J(\vb{x},t) := -\frac{\hbar^2}{2m} \frac{\nabla^2 R_J(\vb{x},t)}{R_J(\vb{x},t)}
\label{eq:quantum_potential}
\end{equation}
is the Bohm (quantum) potential \cite{Bohm1952Part_I}, now a functional of the source through $R_J$. The modified continuity equation for the Born density is
\begin{equation}
\boxed{
\begin{aligned}
\frac{\partial |\psi_J(\vb{x},t)|^2}{\partial t} + \nabla \cdot \bigl[ |\psi_J(\vb{x},t)|^2 \, \vb{v}^{\psi_J}(\vb{x},t) \bigr]
&= \frac{2}{\hbar} \operatorname{Im}\left[ J(\vb{x},t) e^{-\im S_J(\vb{x},t)/\hbar} \right] R_J(\vb{x},t).
\end{aligned}
}
\label{eq:modified_continuity_born}
\end{equation}
The derivation of Eqs.~\eqref{eq:modified_HJ} and \eqref{eq:modified_continuity_born} is given in \ref{app:derivation_HJ_continuity}.

The phase field $S_J$ mixes the real and imaginary parts of the source. Expanding the phase-rotated contributions,
\begin{equation}
\operatorname{Im}\left[ J e^{-\im S_J/\hbar} \right]
= J_{\mathrm{i}} \cos\frac{S_J}{\hbar} - J_{\mathrm{r}} \sin\frac{S_J}{\hbar},
\label{eq:source_imag_expanded}
\end{equation}
\begin{equation}
\operatorname{Re}\left[ J e^{-\im S_J/\hbar} \right]
= J_{\mathrm{r}} \cos\frac{S_J}{\hbar} + J_{\mathrm{i}} \sin\frac{S_J}{\hbar}.
\label{eq:source_real_expanded}
\end{equation}
Unlike a simple additive coupling, the factor $e^{-\im S_J/\hbar}$ rotates the source vector $(J_{\mathrm{r}}, J_{\mathrm{i}})$ in the complex plane, so neither component acts independently. This mixing is a local algebraic feature of the source coupling; it does not by itself imply nonlocality or contextuality, which in dBB theory arise from other aspects of the formalism.

\subsection{Ontological Hierarchy: Particles, Field, and Source}
\label{sec:ontological_hierarchy}

While the continuity equation for the particle distribution $P(\vb{x},t)$ remains unchanged and source-free, the Born density obeys \eqref{eq:modified_continuity_born}, with an explicit source contribution. For clarity, we repeat the equation for $P$ with the source-mediated velocity field made explicit:
\begin{equation}
\frac{\partial P(\vb{x},t)}{\partial t} + \nabla \cdot \bigl[ P(\vb{x},t) \, \vb{v}^{\psi_J}(\vb{x},t) \bigr] = 0.
\label{eq:continuity_P_final}
\end{equation}
This expresses the conservation of particles, independently of the wavefunction dynamics. Integrating over $\mathcal{C}$ with volume measure $\dd \Sigma$, and assuming that the current vanishes sufficiently rapidly at infinity,
\begin{equation}
\frac{\dd}{\dd t} \int_{\mathcal{C}} \dd \Sigma \, P(\vb{x},t) = 0,
\label{eq:probability_conservation_P}
\end{equation}
which holds irrespective of whether $J$ is present. \textit{The particle distribution is governed by a continuity equation that remains source-free, even when the wavefunction dynamics are modified by $J$.} By contrast, integrating \eqref{eq:modified_continuity_born},
\begin{equation}
\frac{\dd}{\dd t} \int_{\mathcal{C}} \dd \Sigma \, |\psi_J(\vb{x},t)|^2
= \frac{2}{\hbar} \int_{\mathcal{C}} \dd \Sigma \, \operatorname{Im}\left[ J(\vb{x},t) e^{-\im S_J(\vb{x},t)/\hbar} \right] R_J(\vb{x},t),
\label{eq:born_normalization_violation_source}
\end{equation}
which is generally nonzero when the source is active. Thus $P$ remains normalized at all times while $|\psi_J|^2$ may change its normalization. This is a direct consequence of the ontological separation between particles and field inherent in the dBB framework.

The key insight is that the dualistic ontology of dBB theory provides a natural resolution to the Schr\"odinger source problem. The source $J$ acts directly only on the pilot-wave $\psi_J$. The distribution $P$ evolves according to its own continuity equation \eqref{eq:continuity_P_final}, which preserves $\int_{\mathcal{C}} \dd \Sigma \, P = 1$ regardless of the source, yet $P$ is indirectly influenced by $J$ through the velocity field $\vb{v}^{\psi_J}$, itself a functional of the source. This gives the following hierarchy:
\begin{itemize}
\item The source $J(\vb{x},t)$ acts directly on $\psi_J(\vb{x},t)$, rendering it a functional of $J$.
\item All field-derived quantities inherit this dependence: $R_J = R[J]$, $S_J = S[J]$, $Q^{\rm B}_J = Q[J]$, and $\vb{v}^{\psi_J} = \vb{v}^{\psi}[J]$.
\item The particle distribution $P(\vb{x},t)$ depends on $J$ only indirectly, via the velocity field $\vb{v}^{\psi_J}$. Its initial configuration $P(\vb{x},0)$ is independent of the source and may be chosen freely, and its evolution equation is source-free and conserves the initial normalization.
\item The Born density $P^{\psi_J}_{\mathrm{B}}(\vb{x},t) := |\psi_J(\vb{x},t)|^2$ is directly prescribed by the source. Even if its initial normalization is unity, this is generally violated while the source is active, so $|\psi_J|^2$ no longer functions as a probability density when $J \neq 0$.
\end{itemize}

\begin{remark}[Phase Symmetry Breaking and the Survival of the Guidance Equation]
\label{rem:phase_symmetry_breaking}
The source $J$ breaks the global $U(1)$ phase symmetry of the source-free theory. Under $\psi_J \to e^{\im\alpha}\psi_J$ (that is, $S_J \to S_J + \hbar\alpha$ with $R_J$ fixed), the density $R_J^2$, the gradient $\nabla S_J$, and the velocity $\mathbf{v} = \nabla S_J/m$ are invariant. The right-hand sides of \eqref{eq:modified_continuity_born} and \eqref{eq:modified_HJ}, $\Sigma_R = (2/\hbar)\operatorname{Im}[J e^{-\im S_J/\hbar}] R_J$ and $\Sigma_S = -(1/R_J)\operatorname{Re}[J e^{-\im S_J/\hbar}]$, acquire a phase factor $e^{-\im\alpha}$ inside the bracket, so the full equations of motion are not $U(1)$-invariant. This is physically natural: the external source provides a phase reference, as an AC voltage source does in circuit theory.

The breaking affects only the right-hand sides of the equations of motion, not the left-hand sides, which retain the canonical structure of the source-free theory. In particular the current $\mathbf{j} = R_J^2 \nabla S_J/m$ is unmodified, so the guidance equation is unaffected: the velocity is determined by $\nabla S_J$ alone, and no additional solenoidal term $\mathbf{W}$ is introduced. A structural justification is given in Remark~\ref{rem:uniqueness_guidance}.
\end{remark}

\begin{remark}[Structural Justification for the Guidance Equation in the Presence of Sources]
\label{rem:uniqueness_guidance}
In the standard framework the guidance equation $\mathbf{v} = \nabla S/m$ is justified by the equivariance of the Born density or simply postulated, as we do here. For $J \neq 0$ the density $R_J^2$ is not conserved, which seemingly allows an arbitrary solenoidal addition, $\mathbf{v} = \nabla S_J/m + \mathbf{W}$. We note two reasons for preferring $\mathbf{W}=0$.

\emph{Kinematic compatibility.} The source enters \eqref{eq:modified_schrodinger} without spatial derivatives. Multiplying the equation by $\psi_J^*$ and subtracting the complex conjugate therefore leaves the spatial current $\mathbf{j} = (\hbar/2\im m)(\psi_J^* \nabla \psi_J - \psi_J \nabla \psi_J^*)$ unchanged, and the source contributes only the term $2\im\operatorname{Im}[\psi_J^* J]$. The continuity equation \eqref{eq:modified_continuity_born} thus keeps the same current on its left-hand side, and $\mathbf{v} := \mathbf{j}/R_J^2 = \nabla S_J/m$ remains the natural identification, which gives no motivation for a solenoidal addition.

\emph{Hamilton--Jacobi structure.} A velocity $\mathbf{v} = \nabla S_J/m + \mathbf{W}$ would require the kinetic term to contain $\nabla S_J \cdot \mathbf{W} + m\mathbf{W}^2/2$, which the Laplacian in \eqref{eq:modified_schrodinger} does not generate. We therefore adopt $\mathbf{W} = 0$ as the simplest choice, which passes smoothly to the conventional theory as $J \to 0$. This is not a proof that $\mathbf{W}$ must vanish. It shows what would be required if a nonzero $\mathbf{W}$ were introduced as part of a different modification of the guidance equation: an externally imposed vector field singles out a preferred direction at each point, whereas the kinetic operator $-(\hbar^2/2m)\nabla^2$ is isotropic and generates only scalar invariants of $\nabla S_J$ and $\nabla R_J$. These terms cannot be moved into $\Sigma_S$ either, since $J$ is independent of $\psi$. Reproducing them would require modifying the Hamiltonian itself, for example by a vector-potential coupling $\nabla S_J \to \nabla S_J - \mathbf{A}$ with $\mathbf{A} = -m\mathbf{W}$, which is a different theory and not a source extension of the present one.
\end{remark}

\subsection{Modified Transport Equation for the Ratio Field $f(\vb{x},t)$}
\label{sec:Modified Transport Equation for the Ratio Field}

In quantum nonequilibrium, where $P(\vb{x},0) \neq |\psi_J(\vb{x},0)|^2$, the source can either amplify the deviation from the Born rule or counteract it. To quantify this, we subtract \eqref{eq:modified_continuity_born} from \eqref{eq:continuity_P_final}, using $f := P/|\psi_J|^2$, to obtain the modified transport equation
\begin{equation}
\boxed{
\begin{aligned}
\frac{D f(\vb{x},t)}{D t}
&:= \frac{\partial f(\vb{x},t)}{\partial t} + \vb{v}^{\psi_J}(\vb{x},t) \cdot \nabla f(\vb{x},t) \\
&= -\frac{2}{\hbar R_J(\vb{x},t)} \operatorname{Im}\left[ J(\vb{x},t) e^{-\im S_J(\vb{x},t)/\hbar} \right] f(\vb{x},t).
\end{aligned}
}
\label{eq:f_modified_final}
\end{equation}
(The full derivation is given in \ref{app:f_derivation}.) When $J = 0$ the right-hand side vanishes and we recover $D f/D t = 0$, the Lagrangian conservation of $f$ that underlies Valentini's subquantum $H$-theorem \cite{Valentini1991a}. When $J \neq 0$, $f$ is no longer conserved along trajectories: the source term acts as a local source or sink for $f$, with the sign set by $\operatorname{Im}[J \exp(-\im S_J/\hbar)]$. If this quantity is positive, $f$ decreases along the trajectory; if negative, $f$ increases. This breaking of the Lagrangian invariance of $f$ at the fine-grained level is a central dynamical fact of the extended theory.

At the trajectory level we apply the method of characteristics \cite{courant_hilbert_1962volue2}. Along a Bohmian trajectory $\vb{q}(t; \vb{q}_0)$ with initial position $\vb{q}_0 := \vb{q}(0)$ we define
\begin{align}
F(t; \vb{q}_0) &:= f(\vb{q}(t; \vb{q}_0), t), \label{eq:F_trajectory} \\
\mathcal{J}(t; \vb{q}_0) &:= J(\vb{q}(t; \vb{q}_0), t), \label{eq:J_trajectory} \\
\mathcal{R}_J(t; \vb{q}_0) &:= R_J(\vb{q}(t; \vb{q}_0), t), \label{eq:R_trajectory} \\
\mathcal{S}_J(t; \vb{q}_0) &:= S_J(\vb{q}(t; \vb{q}_0), t). \label{eq:S_trajectory}
\end{align}
The PDE \eqref{eq:f_modified_final} is then equivalent to the ODE along each characteristic:
\begin{equation}
\boxed{
\frac{\dd F}{\dd t}(t; \vb{q}_0) = -\frac{2}{\hbar \mathcal{R}_J(t; \vb{q}_0)} \operatorname{Im}\left[ \mathcal{J}(t; \vb{q}_0) e^{-\im \mathcal{S}_J(t; \vb{q}_0)/\hbar} \right] F(t; \vb{q}_0),
}
\label{eq:f_ode_characteristics}
\end{equation}
with initial condition $F(0; \vb{q}_0) = f(\vb{q}_0, 0)$. Whether the source drives the system toward or away from $f = 1$ depends on the initial value $F(0)$ and on the subsequent behavior of the source; the system may overshoot, oscillate, or settle into a nonequilibrium steady state. If the source vanishes, $D f/D t = 0$ and we recover Valentini's conservative dynamics.

\section{Driving Quantum Equilibrium: The Role of the Source}
\label{sec:driving_equilibrium}

\subsection{The Source Design Problem}
\label{sec:source_design_problem}

Equation \eqref{eq:f_modified_final} governs the evolution of $f$ along each Bohmian trajectory. A natural question arises: can a suitably chosen source influence the evolution of $f$ in a controlled way? This is interesting for several reasons. First, it opens the door to the potential role of Schr\"odinger source fields in future subquantum technologies. Second, it invites control-theoretic ideas for designing source fields $J$ that steer the system toward desired operational states. In particular, the standard quantum mechanical regime is recovered when $f(\vb{x},t) \to 1$. If admissible sources exist for a given Hamiltonian such that the system approaches quantum equilibrium, one may conjecture that the prevalence of Born-rule statistics---what Valentini has termed ``quantum death'' \cite{Valentini2025beyond_the_quantum}---could be explained by a universal physical source field. Conversely, sources that drive the system away from equilibrium could explain the emergence of nonequilibrium or offer pathways toward its detection.

Suppose we have the ratio field $f(\vb{x},t; \hat H, J)$ \cite{Bohm1953,Valentini1991a,Valentini2025The_trouble}, which depends on the Hamiltonian and the source, and that $f$ is required to follow a desired profile $f_{\rm desired}(\vb{x},t)$. Can a pair $(J, \hat H)$ be found such that
\begin{equation}
f(\vb{x},t; \hat H, J) = f_{\rm desired}(\vb{x},t)
\end{equation}
holds for almost all $\vb{x}$ and $t$?\footnote{To solve \eqref{eq:modified_schrodinger} one needs minimal regularity conditions on $\hat H$ and $J$, introducing a weak interdependence between the two. Throughout this work we assume that all requisite regularity conditions hold.}
We examine a special but important case: exponential relaxation.

\begin{remark}[Forward Dynamical Evolution vs. Inverse Source Design]
\label{rem:forward_vs_inverse}
It is essential to distinguish the forward hierarchy $J \rightarrow \psi_J \rightarrow \vb{v}^{\psi_J} \rightarrow P$ from the inverse source-design approach used in this section. In the forward initial-value problem, $J$ is a prescribed external field and determines $\psi_J$ uniquely. In the inverse problem we prescribe a target dynamics for $f$ and derive the constraint equations that $J$ must satisfy. The required $J$ then emerges as a functional of the state variables ($\psi_J, f$, etc.). This is analogous to antenna theory or optimal control, where a source designed to produce a target field is a functional of that field, without dynamical circularity. The constraint is not a closed-form expression for $J$, and it does not convert $J$ into a dynamical feedback variable: it characterizes the class of admissible sources for which the prescribed $f$-evolution is realized. As shown in Remark~\ref{rem:relaxation_nonuniqueness} and \ref{app:relaxation_nonuniqueness}, the inverse problem is underdetermined: the constraint fixes only one real projection of the complex source, and the family of admissible sources is parameterized by a free real field $\lambda(\vb{x},t)$. The proportional ansatz $J \propto \im\psi_J$ corresponds to the special case $\lambda = 0$ and is a restricted slice of this family, not a generic representative of it.
\end{remark}

\subsection{Exponential Relaxation: Bohmian Trajectory-Level Constraint}
\label{sec:exponential_relaxation_trajectory}

We prescribe the evolution of $f$ along each trajectory by an exponential relaxation with trajectory-dependent parameters:
\begin{equation}
F(t; \vb{q}_0) = 1 + \bigl[ f_0(\vb{q}_0) - 1 \bigr] e^{-t/\tau(\vb{q}_0)},
\label{eq:f_relaxation_general_trajectory}
\end{equation}
where $\tau(\vb{q}_0) > 0$ is the relaxation time. Introduce the \emph{equilibrium deficit}
\begin{equation}
\delta_0(\vb{q}_0) := 1 - f_0(\vb{q}_0),
\label{eq:deficit_trajectory}
\end{equation}
so that
\begin{equation}
F(t; \vb{q}_0) = 1 - \delta_0(\vb{q}_0) e^{-t/\tau(\vb{q}_0)}.
\label{eq:f_relaxation_deficit_trajectory}
\end{equation}
The initial condition for \eqref{eq:f_ode_characteristics} is
\begin{equation}
F(0; \vb{q}_0) = f(\vb{q}_0, 0) =: f_0(\vb{q}_0).
\label{eq:F_initial_condition}
\end{equation}
Substituting \eqref{eq:f_relaxation_general_trajectory} into \eqref{eq:f_ode_characteristics}, and using $\dd F/\dd t = (1 - F)/\tau$, we obtain
\begin{equation}
\frac{1 - F(t; \vb{q}_0)}{\tau(\vb{q}_0)} = -\frac{2}{\hbar \mathcal{R}_J(t; \vb{q}_0)} \operatorname{Im}\left[ \mathcal{J}(t; \vb{q}_0) e^{-\im \mathcal{S}_J(t; \vb{q}_0)/\hbar} \right] F(t; \vb{q}_0).
\label{eq:ode_substituted}
\end{equation}
Solving for the source contribution yields the constraint along each trajectory:
\begin{equation}
\operatorname{Im}\left[ \mathcal{J}(t; \vb{q}_0) e^{-\im \mathcal{S}_J(t; \vb{q}_0)/\hbar} \right]
= -\frac{\hbar \mathcal{R}_J(t; \vb{q}_0)}{2 \tau(\vb{q}_0)} \,
\frac{1 - F(t; \vb{q}_0)}{F(t; \vb{q}_0)}.
\label{eq:source_constraint_general_trajectory}
\end{equation}
Using \eqref{eq:f_relaxation_deficit_trajectory}, this becomes
\begin{equation}
\operatorname{Im}\left[ \mathcal{J}(t; \vb{q}_0) e^{-\im \mathcal{S}_J(t; \vb{q}_0)/\hbar} \right]
= -\frac{\hbar \mathcal{R}_J(t; \vb{q}_0)}{2 \tau(\vb{q}_0)} \,
\frac{\delta_0(\vb{q}_0)\, e^{-t/\tau(\vb{q}_0)}}{1 - \delta_0(\vb{q}_0)\, e^{-t/\tau(\vb{q}_0)}}.
\label{eq:source_explicit_trajectory_deficit}
\end{equation}

\paragraph{From trajectory labels to a field equation.} The functions $f_0(\vb{q}_0)$, $\delta_0(\vb{q}_0)$ and $\tau(\vb{q}_0)$ are Lagrangian: they are attached to trajectories through their initial positions. The source $J(\vb{x},t)$, by contrast, is an Eulerian field. To convert \eqref{eq:source_constraint_general_trajectory} into a field equation we must pull the Lagrangian data back through the flow. Let $\Phi_t:\vb{q}_0 \mapsto \vb{q}(t;\vb{q}_0)$ denote the flow map of \eqref{eq:guidance}, assumed invertible, and define the Eulerian relaxation time
$\tau_{\rm E}(\vb{x},t) := \tau\bigl(\Phi_t^{-1}(\vb{x})\bigr),$
which is the value of $\tau$ carried by the trajectory that passes through $\vb{x}$ at time $t$. It depends on $t$ in general, because the trajectory through $\vb{x}$ changes with time; by construction it is transported by the flow, $\partial_t \tau_{\rm E} + \vb{v}^{\psi_J}\cdot\nabla\tau_{\rm E} = 0$, since $\tau$ is a function of $\vb{q}_0$ alone. The ratio field $f(\vb{x},t)$ on the right-hand side is already Eulerian. The field constraint is then
\begin{equation}
\operatorname{Im}\left[ J(\vb{x},t) e^{-\im S_J(\vb{x},t)/\hbar} \right]
= -\frac{\hbar R_J(\vb{x},t)}{2 \tau_{\rm E}(\vb{x},t)} \,
\frac{1 - f(\vb{x},t)}{f(\vb{x},t)}.
\label{eq:source_constraint_field}
\end{equation}
For uniform $\tau$ one has $\tau_{\rm E} = \tau$ and nothing changes.

Equation \eqref{eq:source_constraint_field} shows that the source contribution required to realize a prescribed relaxation is locally determined by $R_J$, $S_J$, $f$, and the relaxation time $\tau_{\rm E}$. It fixes only one of the two real degrees of freedom of the complex source $J$; the other remains free.

\begin{remark}[Non-Uniqueness of the Relaxation Source and Dual Design Freedom]
\label{rem:relaxation_nonuniqueness}
The constraint \eqref{eq:source_constraint_field} is a single real equation for the two real components of $J$, and is therefore underdetermined. As shown in \ref{app:relaxation_nonuniqueness}, the general solution is parameterized by an arbitrary real scalar field $\lambda(\vb{x},t)$ representing the unconstrained real part of the phase-rotated source. The proportional ansatz $J \propto \im\psi_J$, which in control-theoretic language resembles a linear state-dependent control law \cite{BruntonNoackKoumoutsakos2020,Troltzsch2010}, corresponds to the restricted case $\lambda = 0$. For $\lambda \neq 0$ the source is not of the form $J \propto \im\psi_J$, yet it enforces the same relaxation dynamics. The field $\lambda$ enters the real part of the phase-rotated source, which modifies the effective quantum potential and the Hamilton--Jacobi dynamics \eqref{eq:modified_HJ} without altering the prescribed evolution of $f$. The complex source thus provides two independent design freedoms: the constrained projection fixes the amplitude channel (the relaxation of $f$), while the free field $\lambda$ can be chosen to shape the phase channel (the Hamilton--Jacobi flow and hence the Bohmian trajectories). The underdetermination mirrors the classicalization non-uniqueness of Remark~\ref{rem:classicalization_nonuniqueness}: in both cases the constraint fixes one projection of the phase-rotated source and leaves the orthogonal projection free.
\end{remark}

\subsection{Uniform Relaxation and the Dynamics of Born Integral}
\label{sec:Uniform Relaxation and the Dynamics of Born Integral}

We consider the special case where the initial ratio and the relaxation time are uniform, $f_0(\vb{q}_0) = f_0$ and $\tau(\vb{q}_0) = \tau$ for all $\vb{q}_0 \in \mathcal{C}$, with $f_0 > 0$ and $\tau > 0$. The deficit is then also uniform, $\delta_0 = 1 - f_0$, and \eqref{eq:f_relaxation_general_trajectory} becomes $F(t) = 1 + (f_0 - 1) e^{-t/\tau} = 1 - \delta_0 e^{-t/\tau}$. Substituting into \eqref{eq:source_constraint_field}, we obtain the uniform field constraint
\begin{equation}
\operatorname{Im}\left[ J(\vb{x},t) e^{-\im S_J(\vb{x},t)/\hbar} \right]
= -\frac{\hbar R_J(\vb{x},t)}{2 \tau} \,
\frac{(1 - f_0) e^{-t/\tau}}{1 + (f_0 - 1) e^{-t/\tau}},
\label{eq:source_explicit_uniform_field}
\end{equation}
and along each trajectory the same expression with $R_J \to \mathcal{R}_J$. At $t = 0$,
\begin{equation}
\operatorname{Im}\left[ J(\vb{x},0) e^{-\im S_J(\vb{x},0)/\hbar} \right]
= -\frac{\hbar R_J(\vb{x},0)}{2 \tau} \, \frac{1 - f_0}{f_0},
\label{eq:initial_source_uniform}
\end{equation}
which is finite for $f_0 > 0$. The case $f_0 = 0$ corresponds to an empty initial particle distribution, which cannot be normalized, and the source contribution then diverges as $1/t$ when $t \to 0^+$; we exclude it.

We now examine the evolution of the total Born integral,
\begin{equation}
N_B(t) := \int_{\mathcal{C}} \dd \Sigma \, |\psi_J(\vb{x},t)|^2,
\label{eq:N_B_definition}
\end{equation}
which equals unity for a normalized wavefunction in the source-free limit but is not conserved when a source is present. Using \eqref{eq:modified_continuity_born} together with \eqref{eq:source_constraint_field} we obtain (see \ref{app:proof_NB_evolution})\footnote{The derivation relies on the uniformity of $f$ and $\tau$. In the non-uniform case, $\dd N_B/\dd t = - \int_{\mathcal{C}} \dd \Sigma \, |\psi_J|^2 [1 - f]/[\tau_{\rm E} f]$.}
\begin{equation}
\frac{\dd N_B}{\dd t} = -\frac{1}{\tau} \frac{1 - f(t)}{f(t)} N_B(t).
\label{eq:N_B_evolution}
\end{equation}
Substituting $F(t) = 1 + (f_0 - 1) e^{-t/\tau}$ yields the solution (see \ref{app:derivation_NB_solution})
\begin{equation}
N_B(t) = \frac{N_B(\infty)}{f(t)} = \frac{N_B(\infty)}{1 + (f_0 - 1) e^{-t/\tau}},
\label{eq:N_B_solution_intermediate}
\end{equation}
where $N_B(\infty) := \lim_{t \to \infty} N_B(t)$. In the uniform case, where $f$ is spatially constant, the normalization $\int_{\mathcal{C}} P\,\dd\Sigma = 1$ together with the definition $f = P/|\psi_J|^2$ gives $f(t) N_B(t) = 1$ for all $t$, so $f$ and $N_B$ are reciprocal. At $t = 0$, $f(0) = f_0$, so $N_B(0) = 1/f_0$. As $t \to \infty$, $f(t) \to 1$, so $N_B(\infty) = 1$. In equilibrium ($f_0 = 1$), $N_B(t) = 1$ for all $t$. The uniform case is meant as an illustration of the design procedure of Eq.~\eqref{eq:source_constraint_field}; non-uniform $f_0(\vb{q}_0)$ and $\tau(\vb{q}_0)$ follow the same logic.

\subsection{Interpretation of the Relaxation Source}
\label{sec:Physical Interpretation}

For the source that enforces the uniform relaxation $F(t) = 1 + (f_0 - 1) e^{-t/\tau}$, Eq.~\eqref{eq:source_explicit_uniform_field} gives $\operatorname{Im}[J e^{-\im S_J/\hbar}] < 0$ for $f_0 < 1$, $> 0$ for $f_0 > 1$, and $= 0$ for $f_0 = 1$, where Valentini's conservative dynamics is recovered. By \eqref{eq:f_modified_final}, this sign is the one that makes $f$ increase when $f<1$ and decrease when $f>1$. The conclusions are specific to the relaxation ansatz: the general transport equation \eqref{eq:f_modified_final} imposes no such restriction, and an arbitrary source can drive $f$ either toward or away from equilibrium, depending on the sign of $\operatorname{Im}[J e^{-\im S_J/\hbar}]$ relative to the value of $f$. The opposite scenario, in which the source is engineered to sustain or amplify nonequilibrium, is equally admissible and enters in Section~\ref{sec:quantum_nonequilibrium_source}.
We conclude that suitably chosen source fields can drive $f$ exponentially toward $1$ along every trajectory, with the pair $(f_0, \tau)$, or equivalently $(\delta_0, \tau)$, parameterizing the relaxation. More generally, the source can restore or disrupt the Born rule depending on its configuration.

\section{Quantum Nonequilibrium and Exact Entropy Drift}
\label{sec:quantum_nonequilibrium_source}
\label{sec:entropy_exact}

As an application of the modified dBB dynamics, we investigate how the presence of the scalar field $J$ influences quantum equilibrium and nonequilibrium. In the presence of a source, the modified transport equation \eqref{eq:f_modified_final} implies that Valentini's subquantum $H$-theorem \cite{Valentini1991a} need not be applicable. The $H$-theorem relies on the fact that, for $J = 0$, the ratio $f$ is a Lagrangian invariant along every Bohmian trajectory \cite{Valentini1991a,Bohm1953}; this exact conservation law is the dynamical foundation of the coarse-graining argument. When $J \neq 0$, \eqref{eq:f_modified_final} replaces it with $D f/D t = -(2/\hbar R_J) \operatorname{Im}[ J \exp(-\im S_J/\hbar) ] f$, so the fine-grained entropy is no longer a constant of the motion. In addition, $|\psi_J|^2$ is not conserved when $J \neq 0$ (Eq.~\eqref{eq:born_normalization_violation_source}), so $f$ is no longer a likelihood ratio between two normalized distributions, which independently blocks the relaxation mechanism $P \to |\psi|^2$ of \cite{Bohm1953,Valentini1991a}.
The source thus places the system in a different regime from that considered by Bohm, Valentini, and collaborators: the de Broglie field-particle system becomes \emph{open}, with the source acting as an external agent. A Schr\"odinger source can be used not only to drive the system toward quantum equilibrium, as in Section~\ref{sec:driving_equilibrium}, but also to sustain it away from equilibrium.

To quantify these effects, we use the fine-grained subquantum entropy functional \cite{Valentini1991a}
\begin{equation}
S_{\rm e}(t) := -k_{\rm B} \int_{\mathcal{C}} \dd \Sigma \, P(\vb{x},t) \ln f(\vb{x},t),
\label{eq:entropy_functional}
\end{equation}
where $k_{\rm B}$ is Boltzmann's constant and $f(\vb{x},t) = P(\vb{x},t)/|\psi_J(\vb{x},t)|^2$.\footnote{This functional vanishes identically when $f = 1$ (with $P$ normalized, this requires the Born density to be normalized, $N_B = 1$). When $J \neq 0$ the reference density $|\psi_J|^2$ is generally not normalized; $S_{\rm e}$ is nonetheless well defined whenever the integral converges, and the rate \eqref{eq:entropy_production} below is exact.} Differentiating with respect to time and using \eqref{eq:continuity_P_final} and \eqref{eq:modified_continuity_born}, we obtain (see \ref{app:entropy_production})
\begin{equation}
\boxed{
\frac{\dd S_{\rm e}}{\dd t} = \frac{2 k_{\rm B}}{\hbar} \int_{\mathcal{C}} \dd \Sigma \, \frac{\operatorname{Im}\!\bigl[ J(\vb{x},t) e^{-\im S_J(\vb{x},t)/\hbar} \bigr]}{R_J(\vb{x},t)} \, P(\vb{x},t).
}
\label{eq:entropy_production}
\end{equation}

 We call a nonzero value of $\dd S_{\rm e}/\dd t$ the \emph{entropy drift}. It is new in the extended theory: for $J = 0$ the fine-grained entropy is frozen, $\dd S_{\rm e}/\dd t = 0$, and all change of entropy in Valentini's theory arises only after coarse-graining. With a source, the entropy drifts already at the fine-grained level, and Eq.~\eqref{eq:entropy_production} gives its value exactly, with no coarse-graining and no assumption about the absence of fine-grained microstructure. Equivalently, since $D \ln f/D t = -(2/\hbar R_J)\operatorname{Im}[J e^{-\im S_J/\hbar}]$ along trajectories,
\begin{equation}
\frac{\dd S_{\rm e}}{\dd t} = -k_{\rm B} \int_{\mathcal{C}} \dd \Sigma \, P \, \frac{D \ln f}{D t}.
\label{eq:drift_lagrangian}
\end{equation}
The drift vanishes if $J = 0$ or, more generally, if $\operatorname{Im}[J e^{-\im S_J/\hbar}] = 0$ everywhere on the support of $P$. It has no definite sign: a positive $\operatorname{Im}[J e^{-\im S_J/\hbar}]$ lowers $f$ along a trajectory and contributes positively to the drift, and a negative one does the opposite. These statements are instantaneous. They do not imply that the system reaches $f = 1$, nor that it remains there if it does: a general source may drive $f$ across unity, reverse direction, or produce persistent oscillations. Only specially designed sources, such as the relaxation source of Section~\ref{sec:driving_equilibrium}, in which $J$ is slaved to the instantaneous value of $f$ through \eqref{eq:source_constraint_field}, guarantee monotonic convergence. More generally, a source may be designed to sustain any prescribed value of $f$, including far-from-equilibrium values, by choosing $J$ such that $D f/D t = 0$ at the target $f \neq 1$.

The entropy drift is a diagnostic used in the examples below: it distinguishes sources that leave $f$ conserved along trajectories ($\dd S_{\rm e}/\dd t = 0$) from sources that actively change it.

\section{The Modified Hamilton-Jacobi Equation: Effective Quantum Potential and Classicalization}
\label{sec:real_part_effective_potential}

\subsection{Source-Induced Quantum Potential}
\label{sec:source_induced_potential}

Having established the role of the phase-rotated imaginary projection $\operatorname{Im}[J \exp(-\im S_J/\hbar)]$ in driving the entropy drift and quantum nonequilibrium, we now examine the role of the real projection. As shown in \eqref{eq:modified_HJ}, the source modifies the phase dynamics through the term $(-1/R_J) \operatorname{Re}[ J \exp({-\im S_J/\hbar}) ]$, whose expansion is given in \eqref{eq:source_real_expanded}. This motivates the \emph{source-induced quantum potential}
\begin{equation}
\boxed{
\begin{aligned}
Q^{\rm source}_J(\vb{x},t)
&:= \frac{1}{R_J(\vb{x},t)} \operatorname{Re}\!\bigl[ J(\vb{x},t) e^{-\im S_J(\vb{x},t)/\hbar} \bigr] \\[4pt]
&= \frac{J_{\mathrm{r}}(\vb{x},t)}{R_J(\vb{x},t)} \cos\frac{S_J(\vb{x},t)}{\hbar}
+ \frac{J_{\mathrm{i}}(\vb{x},t)}{R_J(\vb{x},t)} \sin\frac{S_J(\vb{x},t)}{\hbar},
\end{aligned}
}
\label{eq:source_potential}
\end{equation}
so that \eqref{eq:modified_HJ} takes the compact form
\begin{equation}
\frac{\partial S_J}{\partial t} + \frac{(\nabla S_J)^2}{2m} + V + Q^{\rm total}_J = 0,
\label{eq:modified_HJ_compact}
\end{equation}
with the \emph{total quantum potential}
\begin{equation}
\boxed{
Q^{\rm total}_J(\vb{x},t) := Q^{\rm B}_J(\vb{x},t) + Q^{\rm source}_J(\vb{x},t).
}
\label{eq:total_quantum_potential}
\end{equation}
Both $Q^{\rm B}_J$ and $Q^{\rm source}_J$ depend on $J$ through $R_J$ and $S_J$. The distinction lies in their origin and in their source-free limit. The Bohm potential arises from the kinetic term, has the same functional form as in standard dBB theory, and is not directly sourced by $J$. In contrast, $Q^{\rm source}_J$ is directly proportional to the source and vanishes identically when $J = 0$. It is a genuinely new contribution to the phase dynamics, acting as an externally controllable addition to the effective force landscape of the Bohmian particles.

\subsection{Exact Cancellation of the Quantum Potential: Classicalization via the Source}
\label{sec:zero_total_quantum_potential}

Is it possible to choose $J$ such that the total quantum potential vanishes?
\begin{equation}
Q^{\rm total}_J(\vb{x},t) = 0, \qquad \forall \vb{x}, t.
\label{eq:zero_total_quantum_potential}
\end{equation}
If so, \eqref{eq:modified_HJ_compact} reduces to the classical Hamilton-Jacobi equation:
\begin{equation}
\frac{\partial S_J}{\partial t} + \frac{(\nabla S_J)^2}{2m} + V = 0.
\label{eq:classical_HJ}
\end{equation}
In both standard dBB dynamics and classical Hamiltonian mechanics, particles always possess well-defined trajectories \cite{bohm1987quantum_implications,bohm1993the_undivided_universe,durrr2009bohmian_mechanics,holland1995the_quantum_theory_of_motion,Valentini1991a}. What distinguishes \eqref{eq:zero_total_quantum_potential} is the statistical phenomenology of the ensemble. When $Q^{\rm total}_J = 0$, the velocity field $\vb{v}^{\psi_J} = \nabla S_J/m$ is governed by the classical Hamilton-Jacobi equation, the particle ensemble obeys the classical Liouville equation, and the interference phenomena that in dBB theory arise from the amplitude dependence of the velocity field are suppressed at the level of particle statistics.\footnote{For how, in standard dBB theory, individual trajectories in the double-slit experiment reproduce the interference pattern in the ensemble, see \cite{bohm1993the_undivided_universe}. When $Q^{\rm total}_J = 0$ the guiding wave no longer imprints its amplitude structure onto the particle flow.}

Using \eqref{eq:total_quantum_potential}, the condition becomes $Q^{\rm source}_J = -Q^{\rm B}_J$, i.e.
\begin{equation}
\frac{J_{\mathrm{r}}(\vb{x},t)}{R_J(\vb{x},t)} \cos\frac{S_J(\vb{x},t)}{\hbar}
+ \frac{J_{\mathrm{i}}(\vb{x},t)}{R_J(\vb{x},t)} \sin\frac{S_J(\vb{x},t)}{\hbar}
= -Q^{\rm B}_J(\vb{x},t).
\label{eq:zero_total_constraint}
\end{equation}
This is a single real equation for $J_{\mathrm{r}}$ and $J_{\mathrm{i}}$. A natural solution aligns the source vector with the phase factor:
\begin{equation}
J_{\mathrm{r}}(\vb{x},t) = -R_J(\vb{x},t) \, Q^{\rm B}_J(\vb{x},t) \cos\frac{S_J(\vb{x},t)}{\hbar},
\label{eq:Jr_zero_total}
\end{equation}
\begin{equation}
J_{\mathrm{i}}(\vb{x},t) = -R_J(\vb{x},t) \, Q^{\rm B}_J(\vb{x},t) \sin\frac{S_J(\vb{x},t)}{\hbar}.
\label{eq:Ji_zero_total}
\end{equation}
Substituting into \eqref{eq:zero_total_constraint} gives $-Q^{\rm B}_J [\cos^2(S_J/\hbar) + \sin^2(S_J/\hbar)] = -Q^{\rm B}_J$, an identity. Combining the components, the special source takes the compact form
\begin{equation}
J(\vb{x},t) = -Q^{\rm B}_J(\vb{x},t) \, \psi_J(\vb{x},t), \qquad \forall \vb{x}, t.
\label{eq:J_compact_zero_total}
\end{equation}
This relation is a \emph{constraint} on the source, not a closed-form expression. Since $Q^{\rm B}_J$ and $\psi_J$ depend on $J$ through \eqref{eq:modified_schrodinger}, \eqref{eq:J_compact_zero_total} is an implicit self-consistency condition. The modified Schr\"odinger equation becomes a nonlinear equation:
\begin{equation}
\im \hbar \frac{\partial \psi_J}{\partial t}
= -\frac{\hbar^2}{2m} \nabla^2 \psi_J + V \psi_J - Q^{\rm B}_J \psi_J,
\label{eq:nonlinear_schrodinger}
\end{equation}
where $Q^{\rm B}_J$ depends nonlinearly and nonlocally on $\psi_J$ via \eqref{eq:quantum_potential}.\footnote{We do not establish here global existence and uniqueness for this nonlinear equation; the classicalization constraint is treated as a formal condition, and its solutions are assumed to exist for the times of interest. A fully rigorous and comprehensive treatment is left to future work.}

\begin{remark}[Non-Uniqueness of Classicalization]
\label{rem:classicalization_nonuniqueness}
The condition $Q^{\rm total}_J = 0$ requires $\operatorname{Re}[J e^{-\im S_J/\hbar}] = -Q^{\rm B}_J R_J$ and places no constraint on the imaginary part. The specific solution $J = -Q^{\rm B}_J \psi_J$ is the restricted case $\operatorname{Im}[J e^{-\im S_J/\hbar}] = 0$, the reverse Madelung slice (see, e.g., \cite{holland1995the_quantum_theory_of_motion}); because the imaginary part vanishes, the entropy drift is zero there. As shown in \ref{app:alternative_classicalization}, the general classicalizing source is parameterized by an arbitrary real scalar field $\mu(\vb{x},t)$ for the unconstrained imaginary part. For any $\mu \neq 0$ the source lies outside the reverse Madelung slice, yet it still yields trajectories governed by the classical Hamilton--Jacobi guidance law.
\end{remark}

\subsection{Subquantum Classical Regime: The Reverse Madelung Slice}
\label{sec:subquantum_classical_regime}

With $Q^{\rm total}_J = 0$, the guidance equation retains its form $\dd \vb{q}/\dd t = (1/m) \nabla S_J(\vb{q}, t)$, but now $S_J$ satisfies \eqref{eq:classical_HJ}. The Bohmian particles traverse trajectories dictated solely by the classical potential $V$. The quantum potential, which in the original dBB theory is the conduit through which the wavefunction's non-classical features influence particle motion \cite{bohm1993the_undivided_universe}, has been neutralized by the source.
The mechanism acts on the \emph{guidance dynamics} rather than on the wavefunction: the guiding wavefunction $\psi_J$ continues to be governed by the nonlinear equation \eqref{eq:nonlinear_schrodinger}. Within this framework ``classical particle trajectories'' and ``classical wavefunction'' are not equivalent notions: the trajectories become classical while the guiding field retains a nontrivial amplitude and phase structure.\footnote{Whether the nonlinearity of \eqref{eq:nonlinear_schrodinger} permits or suppresses specific quantum features such as long-range interference or superposition is a dynamical question depending on the initial conditions. Any such structure that survives in $\psi_J$ no longer imprints itself on the particle motion, because the quantum potential has been eliminated.} This decoupling has no analogue in standard quantum mechanics, where classical motion emerges only through decoherence or the $\hbar \to 0$ limit, in which the Born rule is restored.

What of the entropy drift in this regime? Substituting \eqref{eq:Jr_zero_total}--\eqref{eq:Ji_zero_total} into \eqref{eq:source_imag_expanded},
$\operatorname{Im}[J \exp(-\im S_J/\hbar)] = J_{\mathrm{i}} \cos(S_J/\hbar) - J_{\mathrm{r}} \sin(S_J/\hbar) = 0$.
The source vector $(J_{\mathrm{r}}, J_{\mathrm{i}})$ is parallel (or anti-parallel, depending on the sign of $Q^{\rm B}_J$) to the phase direction $(\cos(S_J/\hbar), \sin(S_J/\hbar))$, which is the projection entering $\operatorname{Re}[J \exp(-\im S_J/\hbar)]$ and hence $Q^{\rm source}_J$. The orthogonal direction $(-\sin(S_J/\hbar), \cos(S_J/\hbar))$, which defines the imaginary projection responsible for the drift via \eqref{eq:entropy_production}, receives no projection. The cancellation is therefore selective: the source acts along the phase direction to neutralize the Bohm potential and leaves the amplitude channel inert.
From \eqref{eq:entropy_production} we obtain
$\dd S_{\rm e}/\dd t = 0$.
The meaning of this vanishing differs from the source-free case. For $J = 0$ it reflects the absence of any external influence on the ensemble. Here $J$ is nonzero and actively modifies the Hamilton-Jacobi flow, yet the source is tuned so that it contributes nothing to the amplitude channel.

With $\operatorname{Im}[J \exp(-\im S_J/\hbar)] = 0$, \eqref{eq:f_modified_final} reduces to $D f/D t = 0$, and $f$ is a Lagrangian invariant, as in source-free dBB dynamics. This is a special feature of the reverse Madelung slice ($\mu = 0$). For the general route ($\mu \neq 0$) of Remark~\ref{rem:classicalization_nonuniqueness} and \ref{app:alternative_classicalization}, $f$ is not conserved and the entropy drift is nonzero. Both equilibrium ($f = 1$) and nonequilibrium ($f \neq 1$) ensembles are possible and persist for the slice $\mu = 0$. In the nonequilibrium case one obtains a non-Born ensemble whose trajectories are strictly classical, governed by $V$ alone, while $\psi_J$ obeys \eqref{eq:nonlinear_schrodinger}. The classicality of the trajectories and the complexity of the guiding field are independent in this regime.

\subsection{Non-Madelung Classicalization and the Entropy Drift}
\label{sec:open_system_classicalization}

The reverse Madelung slice is the case $\mu = 0$ of the general classicalization family. It describes classical motion with a vanishing entropy drift. The members with $\mu(\vb{x},t) \neq 0$ define a further regime, summarized here, with details in \ref{app:alternative_classicalization}.
For $\mu \neq 0$ the classicalization constraint still holds, so the total quantum potential vanishes and the trajectories are classical. However, $f$ is no longer a Lagrangian invariant: it obeys $D f/D t = -(2/\hbar R_J)\mu f$. The entropy drift $\dd S_{\rm e}/\dd t$ then contains a contribution from the change of the norm $N_B$, and the contribution that remains after removing it can have either sign, depending on how the source parameters are distributed under $P$ and under the Born density. The classicality of the trajectories and the constancy of $f$ are thus decoupled phenomena: for $\mu = 0$ the trajectories are classical and $f$ is conserved, for $\mu \neq 0$ the trajectories are still classical but $f$ evolves and the entropy drifts. The reverse Madelung transformation is a mathematical mapping lying on the $f$-conserving slice of the family, and the extended theory admits a richer set of behaviors than that slice alone, which may be of interest in programs where quantization is approached from a subquantum level. Whether this larger set is physically realizable, and how it might be probed, are questions for future work.

\section{Concluding Remarks: The Ontological Status of the Source Field}
\label{sec:outlook}

The present work has shown that a source term can be consistently incorporated into the de Broglie-Bohm pilot-wave framework, and that doing so opens a landscape of dynamical possibilities---relaxation to equilibrium, sustained nonequilibrium, and source-induced classicalization. The extended framework treats standard quantum mechanics as the quantum-equilibrium regime of a broader subquantum theory. The broader scope relieves the pressure to derive quantum equilibrium as an inescapable global attractor; it shifts the focus to understanding how the source field is distributed in nature or how it might be deliberately controlled.

\subsection{The Probability Objection and Its Resolution}
\label{sec:probability_objection}

The traditional objection to source terms in the Schr\"odinger equation---that they violate probability conservation---is neutralized within the dualistic ontology of dBB theory. Probability is vested in the particle distribution $P(\vb{x},t)$, whose continuity equation \eqref{eq:continuity_P_final} is kinematic in origin and strictly source-free irrespective of $J$. The Born density $|\psi_J|^2$ need not be conserved, because it no longer functions as a probability measure. The objection rests on a conflation of the Born density with the physical distribution of particles, which the dBB ontology rejects. The source $J$ can then be interpreted as a physical agent capable of driving the particle ensemble into, out of, or independently of quantum equilibrium, without violating conservation of total probability. The framework extends the subquantum program of Valentini and collaborators \cite{Valentini1991a,Valentini1991b} from the analysis of passive relaxation to actively controlled nonequilibrium states.

\subsection{The Extended Ontology: Three Fundamental Fields}
\label{sec:extended_ontology}

The extended ontology has three fundamental fields: the de Broglie pilot-wave field $\psi(\vb{x},t)$, a complex scalar field on configuration space that guides the particles; the particle distribution $P(\vb{x},t)$, a real field encoding the statistical density of the ensemble; and the Schr\"odinger source field $J(\vb{x},t)$, a complex scalar field that generates the pilot wave through \eqref{eq:modified_schrodinger}.

In the dualistic ontology of the source-free theory the joint specification of the pair $(\psi, P)$ constitutes the complete physical description. The coupling between $\psi$ and $P$ is there strictly minimal: $P$ is merely advected along the flow generated by $\vb{v}^\psi$, and $f$ is a Lagrangian invariant. This minimal kinematic coupling is what opens the door to the extension. The condition $D f/D t = 0$ is structurally analogous to Newton's first law: in the absence of external influences, a first-order relational quantity built from the primitive fields remains constant along the flow. The natural extension seeks a subquantum analogue of the second law, in which the material derivative of $f$ is driven by an external agent. The source $J$ does this through \eqref{eq:f_modified_final}: it breaks the kinematic lock between $\psi$ and $P$, and the entropy drift \eqref{eq:entropy_production} measures the resulting change in the relation between them.

In the language of \emph{primitive ontology}, one may view the extended theory as consisting of $\psi$, $P$, and $J$ as primitive structures, none of which is reducible to the others. $P$ being a real field requires some care, since it is a probability distribution; for our purposes it suffices to treat the theory as statistical (an ensemble formalism on configuration space), and the source $J$ as ontologically distinct from the Hamiltonian $\hat H$: the system is specified by $(\hat H, J)$ and its state by $(\psi, P)$.\footnote{For discussion of whether Bohm himself was a determinist or an indeterminist, see \cite{oldofredi2025pilotwave}.}
The causal structure is hierarchical: $J \rightarrow \psi \rightarrow \vb{v}^{\psi_J} \rightarrow P$. Given $J$ and the initial conditions, $\psi$ and $P$ are fully dynamical. The source $J$, however, is treated as \emph{external}: its dynamics is unspecified and the evolution of $\psi$ and $P$ does not react back upon it. In this respect the theory follows classical field theory, where the charge-current distribution in electrodynamics or the energy-momentum tensor in general relativity are prescribed functions determining the dynamical fields.\footnote{For a defense of the view that the ultimate sources of classical fields are continuous fields rather than discrete particles, see \cite{Wald2022electromagnetism}.} Whether the dynamics of $J$ is classical, quantum, or subquantum we leave open.\footnote{For a response to the critique that pilot-wave theories lack back-reaction of particles on the guiding wave, see \cite{Valentini2025The_trouble}. The treatment of $J$ as external is a feature of the current stage of development.}

\subsection{Ontological versus Phenomenological Interpretations}
\label{sec:ontological_vs_phenomenological}

Two broad interpretative stances are available for $J$, which we term the \emph{ontological} (strong) and the \emph{phenomenological} (weak).

In the ontological interpretation, $J$ is a fundamental physical field on an equal footing with $\psi$. Just as a charge distribution produces an electromagnetic field, $J$ produces $\psi$ through the modified Schr\"odinger equation. The universal prevalence of Born-rule statistics---Valentini's ``quantum death'' \cite{Valentini2025beyond_the_quantum}---then reflects the action of a cosmological source field, and the relaxation dynamics of Section~\ref{sec:driving_equilibrium} describe how a generic initial nonequilibrium distribution is brought into conformity with the Born rule. Localized departures from equilibrium could arise wherever $J$ deviates from its cosmological average, providing a potential signature of subquantum physics.

In the phenomenological interpretation, $J$ is not fundamental but encodes the effective influence of degrees of freedom not included in the Hamiltonian:
\begin{enumerate}
\item \emph{Environmental coupling.} Tracing out an external system with many degrees of freedom can generate an effective source term in the reduced Schr\"odinger equation.\footnote{A familiar example is the derivation of the Gorini-Kossakowski-Sudarshan-Lindblad (GKSL) master equation \cite{GKSL976GKSL_master_equation,Lindblad1976ME}: the reduced dynamics is non-unitary as a consequence of discarding information about the environment. The inhomogeneous term $J$ is structurally analogous, encoding the influence of degrees of freedom external to the field-particle system.}
\item \emph{Wavefunction collapse.} A deterministic source term may be viewed as the mean-field limit of stochastic collapse dynamics, capturing effective localization without specifying the underlying noise process.
\item \emph{Subquantum corrections.} $J$ may be the leading-order effective correction from a deeper theory, in the same spirit in which the Navier-Stokes equation emerges from Boltzmann kinetics or effective field theories encode unknown ultraviolet physics.\footnote{The modified equation \eqref{eq:modified_schrodinger} remains linear; it is inhomogeneous, with general solution the sum of the unitary evolution of the initial state and a particular integral over $J$. Whether the non-conservation of the Born norm signals a breakdown of unitarity or merely the openness of the subsystem depends on one's interpretative stance.}
\item \emph{Control-theoretic device.} $J$ can be treated as an external control parameter chosen to steer the particle distribution toward a desired target, as in the relaxation and classicalization constructions of Sections~\ref{sec:driving_equilibrium} and~\ref{sec:real_part_effective_potential}.
\end{enumerate}
A natural extension of the phenomenological view is to promote $J$ to a stochastic field and study the resulting stochastic version of the theory; its ensemble-averaged dynamics may then display irreversible behavior. We leave this to future work.

\subsection{Why the Born Rule?}
\label{sec:why_born_rule}

If the source-extended framework allows for persistent nonequilibrium, why does the observable world appear to obey the Born rule? The source-free answer, discussed in \cite{Valentini2025beyond_the_quantum}, is that relaxation to equilibrium is a generic consequence of coarse-graining over primordial nonequilibrium. We emphasize that the goal of the present paper is not to recover standard quantum mechanics by modifying the Schr\"odinger equation. The extended theory is a covering subquantum theory, and standard quantum mechanics is its special case $J = 0$ with $f_0 = 1$. The question is therefore in which regimes of the extended theory Born-like statistics are recovered, and how this recovery differs from standard quantum mechanics. We distinguish three tiers. 

\emph{Tier 1: the source-free route.} For $J = 0$ and $f_0 = 1$, Bohm's equivariance argument \cite{Bohm1952Part_I,Bohm1952Part_II} reproduces standard quantum mechanics exactly, and, for an initial mismatch, relaxation by coarse-graining is the route discussed above.

\emph{Tier 2: localized sources.} Suppose first that $J$ is localized in time, vanishing outside an interval $[t_1, t_2]$. For $t > t_2$ the equation is homogeneous and $\psi_J$ evolves unitarily from its value at $t_2$, so that $D f/D t = 0$ again. The dynamics is then exactly source-free, with initial data $\psi_J(t_2)$ and $f(t_2)$; if $f(t_2) = 1$, standard dBB dynamics with Born statistics follows. Suppose now that $J$ is localized in configuration space, supported in a region $\Omega_J$ away from the laboratory region $\Omega_{\rm lab}$. Since the Schr\"odinger equation has no finite propagation cone, $\psi_J$ in $\Omega_{\rm lab}$ is influenced by $J$, but the influence is small when the Duhamel integral $\hbar^{-1}\int \|J\|\,\dd t$ is small there or when the wave packet has left $\Omega_J$. Recovery of Born-like statistics in $\Omega_{\rm lab}$ is then \emph{approximate}, up to tails, and the standard route to the Born rule applies with the corresponding accuracy. Where $J=0$ the equation is locally homogeneous and $D f/D t = 0$ exactly, in analogy with the incident field in electromagnetic scattering, which satisfies the homogeneous Helmholtz equation inside a source-free region \cite{felsen1994radiation,Chew_1990,Van_Bladel_2007}; the exterior source acts only through the field carried into the region.

\emph{Tier 3: phase-locked sources.} If the imaginary projection vanishes, $\operatorname{Im}[J e^{-\im S_J/\hbar}] = 0$, then $|\psi_J|^2$ is conserved, $f$ is a Lagrangian invariant, and the entropy drift vanishes. Consequently, if Born-like statistics $P = |\psi_J|^2$ hold at the initial time (with the Born density normalized), they are preserved for all times, even for a substantial source amplitude; if the initial ratio $f$ differs from unity, a phase-locked source preserves that nonequilibrium rather than eliminating it. Writing $J = |J|e^{\im\phi_J}$, the condition holds when $\phi_J = S_J/\hbar + n\pi$, $n \in \mathbb{Z}$. The source remains externally prescribed: the condition constrains admissible $J$ jointly with the solution $\psi_J$ and does not make $J$ a functional of $\psi_J$. The real projection need not vanish, and then $\psi_J$ differs from the wavefunction $\psi_0$ of standard quantum mechanics with the same Hamiltonian and initial data, because the source-induced quantum potential is amplitude-dependent and does not reduce to a shift of $V$. Born-like statistics are thus \emph{not} equivalent to recovering standard quantum mechanics: the same position statistics can be generated by a different guiding wave. This is not a failure of the theory to recover quantum mechanics, since recovering it was never the aim; it shows the richness of the covering theory. It also makes the two cases experimentally distinguishable in principle. If position measurements yield Born statistics for a guiding wave $\psi_J$ whereas other measurements (for instance, interference experiments with superposed inputs) require a wavefunction different from the one predicted by $J = 0$, this suggests the presence of an actual source.\footnote{The special case $J = w_0\psi_J$ with $w_0$ real and constant is unitarily equivalent to a shift $V \to V+w_0$; the test is then whether the data require $V+w_0$ rather than $V$. The general external source satisfying the phase-locking condition is not of this form and is not necessarily reducible to a potential shift. A complex source violates phase-locking and drives the entropy drift.} The simultaneous vanishing of both projections requires $J = 0$.

\emph{Cosmological sources.} More speculatively, $J$ may be identified with, or coupled to, scalar fields already posited in cosmology, such as the inflaton or ultralight dark-sector scalars. The Born rule would then be a local approximation, valid where such a field has decayed or has settled into a phase-locked configuration, while substantial deviations $f \neq 1$ would be expected where it remains active, as in the early universe or near compact objects.

These scenarios cannot be evaluated without empirical input. They motivate tests of the Born rule beyond the laboratory, in astrophysical and cosmological settings and in high-precision terrestrial experiments \cite{Valentini2025beyond_the_quantum}, and a search for regimes in which the source projections remain dynamically active, with sustained far-from-equilibrium states and a nonzero exact entropy drift.

\appendix

\section{Relationship to Other Source-Modified Schr\"odinger Equations}
\label{app:source_comparisons}

The introduction of a source term into the Schr\"odinger equation invites comparison with existing literature on inhomogeneous modifications of quantum wave equations. Their foundational motivations, ontological commitments, and mathematical structures differ from those of the extended dBB framework. We compare with three research programs.

\textit{1. Holland's double-solution program.}
Holland \cite{Holland2019} constructs a unified field theory of wave and particle by introducing an auxiliary field $u$ satisfying an inhomogeneous Schr\"odinger-type equation, whose source is the Bohmian particle itself, represented as a delta-function singularity at $\vb{q}_p(t)$. The pilot wave $\psi$ remains source-free throughout. Holland's source is thus localized at the particle position, in contrast with the continuous source $J(\vb{x},t)$ on configuration space used here. Holland's objective is to derive the guidance equation from Noether symmetries within a monist ontology, whereas we treat the guidance equation as an axiom and pursue a field-theoretic completion of the dualistic ontology.

\textit{2. Particle-centered action-reaction models.}
Sutherland \cite{Sutherland2019a,Sutherland2019b} and Darrow and Bush \cite{Darrow2024} restore action-reaction symmetry within pilot-wave theory by letting the particle act back on the guiding wave through a delta-function source $\delta(\vb{x} - \vb{q}_p(t))$. These approaches lead to nonlinear, singular dynamics requiring regularization or averaging to recover the linear Schr\"odinger equation. In our framework the source $J(\vb{x},t)$ is decoupled from the particle's instantaneous position; it acts exclusively on $\psi_J$, while $P$ evolves by a source-free continuity equation. The programs are complementary.

\textit{3. Phenomenological and effective source mechanisms.}
Effective source terms arise from complex potentials \cite{Bender2007}, from the no-jump evolution of quantum trajectories (generated by $H_{\rm eff} = H - (\im\hbar/2)\sum_k L_k^\dagger L_k$ \cite{GKSL976GKSL_master_equation}), and from dynamical collapse models, such as the Ghirardi-Rimini-Weber (GRW) theory \cite{GRW1986} and continuous spontaneous localization (CSL) \cite{Pearle1989}, which add stochastic, state-dependent terms to solve the measurement problem (see \cite{BassiGhirardi2003,Bassietal2013,AdlerBassi2009,PearleValentini2006} for reviews). These sources are state-dependent and phenomenological, introduced to model environmental coupling, measurement, or collapse. Our source $J(\vb{x},t)$ is externally specified, continuous on configuration space, and not required to be a functional of $\psi$; it does not modify the spectral properties of the Hamiltonian, and the particle distribution $P$ remains normalized.

The contrast with GRW and CSL is sharper still. In those models the modification of the Schr\"odinger equation is designed to reproduce the Born rule (and collapse) from a state-dependent stochastic dynamics, whereas here the source is deterministic for a given $J$, is not designed to recover Born statistics, and has implications beyond the quantum regime, since it governs departures from Born-like statistics as well as their recovery. In the present theory, standard quantum mechanics is the special case $J = 0$, not the target of the construction.

A fuller discussion of the interpretative options is given in Section~\ref{sec:ontological_vs_phenomenological}.

\section{Derivation of the Modified Hamilton-Jacobi Equation \eqref{eq:modified_HJ} and Continuity Equation \eqref{eq:modified_continuity_born}}
\label{app:derivation_HJ_continuity}

Starting from \eqref{eq:modified_schrodinger} with $J = J_{\mathrm{r}} + \im J_{\mathrm{i}}$ and $\psi_J = R_J \exp(\im S_J/\hbar)$, we separate real and imaginary parts. The time derivative of $\psi_J$ is
\begin{equation}
\frac{\partial \psi_J}{\partial t}
= \left( \frac{\partial R_J}{\partial t} + \frac{\im}{\hbar} R_J \frac{\partial S_J}{\partial t} \right) e^{\im S_J/\hbar}.
\end{equation}
Using $\nabla^2 (fg) = f \nabla^2 g + 2 \nabla f \cdot \nabla g + g \nabla^2 f$, the Laplacian is
\begin{equation}
\nabla^2 \psi_J
= \left[ \nabla^2 R_J + \frac{2\im}{\hbar} \nabla R_J \cdot \nabla S_J + \frac{\im}{\hbar} R_J \nabla^2 S_J - \frac{1}{\hbar^2} R_J (\nabla S_J)^2 \right] e^{\im S_J/\hbar}.
\end{equation}
Substituting into \eqref{eq:modified_schrodinger} and dividing by $e^{\im S_J/\hbar}$,
\begin{equation}
\begin{aligned}
\im \hbar \left( \frac{\partial R_J}{\partial t} + \frac{\im}{\hbar} R_J \frac{\partial S_J}{\partial t} \right)
&= V R_J + (J_{\mathrm{r}} + \im J_{\mathrm{i}}) e^{-\im S_J/\hbar} \\
&\quad - \frac{\hbar^2}{2m} \left( \nabla^2 R_J + \frac{2\im}{\hbar} \nabla R_J \cdot \nabla S_J + \frac{\im}{\hbar} R_J \nabla^2 S_J - \frac{1}{\hbar^2} R_J (\nabla S_J)^2 \right).
\end{aligned}
\end{equation}
The source term expands as
\begin{equation}
(J_{\mathrm{r}} + \im J_{\mathrm{i}}) e^{-\im S_J/\hbar}
= (J_{\mathrm{r}} + \im J_{\mathrm{i}})\left(\cos\frac{S_J}{\hbar} - \im \sin\frac{S_J}{\hbar}\right).
\end{equation}
Collecting the real terms,
\begin{equation}
- R_J \frac{\partial S_J}{\partial t}
= -\frac{\hbar^2}{2m} \nabla^2 R_J + \frac{1}{2m} R_J (\nabla S_J)^2 + V R_J + J_{\mathrm{r}} \cos\frac{S_J}{\hbar} + J_{\mathrm{i}} \sin\frac{S_J}{\hbar}.
\end{equation}
Multiplying by $-1$ and dividing by $R_J$,
\begin{equation}
\frac{\partial S_J}{\partial t} + \frac{(\nabla S_J)^2}{2m} + V - \frac{\hbar^2}{2m} \frac{\nabla^2 R_J}{R_J}
= -\frac{J_{\mathrm{r}}}{R_J} \cos\frac{S_J}{\hbar} - \frac{J_{\mathrm{i}}}{R_J} \sin\frac{S_J}{\hbar},
\end{equation}
which with \eqref{eq:quantum_potential} is \eqref{eq:modified_HJ}. Collecting the imaginary terms,
\begin{equation}
\hbar \frac{\partial R_J}{\partial t}
= -\frac{\hbar}{m} \nabla R_J \cdot \nabla S_J - \frac{\hbar}{2m} R_J \nabla^2 S_J + J_{\mathrm{i}} \cos\frac{S_J}{\hbar} - J_{\mathrm{r}} \sin\frac{S_J}{\hbar}.
\end{equation}
Dividing by $\hbar$,
\begin{equation}
\frac{\partial R_J}{\partial t}
= -\frac{1}{m} \left( \nabla R_J \cdot \nabla S_J + \frac{1}{2} R_J \nabla^2 S_J \right) + \frac{J_{\mathrm{i}}}{\hbar} \cos\frac{S_J}{\hbar} - \frac{J_{\mathrm{r}}}{\hbar} \sin\frac{S_J}{\hbar}.
\end{equation}
Since $\nabla \cdot (R_J^2 \nabla S_J) = 2 R_J \nabla R_J \cdot \nabla S_J + R_J^2 \nabla^2 S_J$,
\begin{equation}
\nabla R_J \cdot \nabla S_J + \frac{1}{2} R_J \nabla^2 S_J = \frac{1}{2 R_J} \nabla \cdot (R_J^2 \nabla S_J),
\end{equation}
so
\begin{equation}
\frac{\partial R_J}{\partial t}
= -\frac{1}{2m R_J} \nabla \cdot (R_J^2 \nabla S_J) + \frac{J_{\mathrm{i}}}{\hbar} \cos\frac{S_J}{\hbar} - \frac{J_{\mathrm{r}}}{\hbar} \sin\frac{S_J}{\hbar}.
\end{equation}
Multiplying by $2 R_J$ and using $2 R_J \partial_t R_J = \partial_t |\psi_J|^2$ and $\vb{v}^{\psi_J} = \nabla S_J / m$,
\begin{equation}
\frac{\partial |\psi_J|^2}{\partial t} + \nabla \cdot \left( |\psi_J|^2 \vb{v}^{\psi_J} \right)
= \frac{2}{\hbar} \operatorname{Im}\left[ J e^{-\im S_J/\hbar} \right] R_J,
\end{equation}
which is \eqref{eq:modified_continuity_born}.

\section{Derivation of the Modified Transport Equation for $f$, Eq.~\eqref{eq:f_modified_final}}
\label{app:f_derivation}

Set $P = f |\psi_J|^2$ in \eqref{eq:continuity_P_final}:
\begin{equation}
\frac{\partial}{\partial t} \bigl( f |\psi_J|^2 \bigr) + \nabla \cdot \bigl( f |\psi_J|^2 \vb{v}^{\psi_J} \bigr) = 0.
\end{equation}
Expanding with the product rule and rearranging to isolate the material derivative of $f$,
\begin{equation}
|\psi_J|^2 \left( \frac{\partial f}{\partial t} + \vb{v}^{\psi_J} \cdot \nabla f \right)
+ f \left[ \frac{\partial |\psi_J|^2}{\partial t} + \nabla \cdot \bigl( |\psi_J|^2 \vb{v}^{\psi_J} \bigr) \right] = 0.
\end{equation}
Using \eqref{eq:modified_continuity_born} for the bracket,
\begin{equation}
|\psi_J|^2 \left( \frac{\partial f}{\partial t} + \vb{v}^{\psi_J} \cdot \nabla f \right)
+ f \left( \frac{2}{\hbar} \operatorname{Im}\left[ J e^{-\im S_J/\hbar} \right] R_J \right) = 0.
\end{equation}
Dividing by $|\psi_J|^2 = R_J^2$,
\begin{equation}
\frac{\partial f}{\partial t} + \vb{v}^{\psi_J} \cdot \nabla f = -\frac{2}{\hbar R_J} \operatorname{Im}\left[ J e^{-\im S_J/\hbar} \right] f,
\end{equation}
which is \eqref{eq:f_modified_final}.

\section{Derivation of the entropy drift \eqref{eq:entropy_production}}
\label{app:entropy_production}

Differentiating \eqref{eq:entropy_functional} in time,
\begin{equation}
\frac{\dd S_{\rm e}}{\dd t} = -k_{\rm B} \int_{\mathcal{C}} \dd \Sigma \,
\left[ \frac{\partial P}{\partial t} \ln f + P \frac{1}{f} \frac{\partial f}{\partial t} \right].
\end{equation}
By \eqref{eq:continuity_P_final}, $\partial P/\partial t = -\nabla \cdot ( P \vb{v}^{\psi_J} )$. From the transport equation \eqref{eq:f_modified_final} (Appendix~\ref{app:f_derivation}),
\begin{equation}
\frac{\partial f}{\partial t} = - \vb{v}^{\psi_J} \cdot \nabla f - \frac{2 \operatorname{Im}\!\bigl[ J e^{-\im S_J/\hbar} \bigr]}{\hbar R_J} f.
\end{equation}
Substituting,
\begin{equation}
\frac{\dd S_{\rm e}}{\dd t} = -k_{\rm B} \int_{\mathcal{C}} \dd \Sigma \,
\left[ -\nabla \cdot \left( P \vb{v}^{\psi_J} \right) \ln f - P \vb{v}^{\psi_J} \cdot \nabla \ln f - \frac{2 \operatorname{Im}\!\bigl[ J e^{-\im S_J/\hbar} \bigr]}{\hbar R_J} P \right].
\end{equation}
The first two terms combine to a total divergence,
\begin{equation}
-\nabla \cdot \left( P \vb{v}^{\psi_J} \right) \ln f - P \vb{v}^{\psi_J} \cdot \nabla \ln f = -\nabla \cdot \left( P \vb{v}^{\psi_J} \ln f \right).
\end{equation}
Assuming the boundary term vanishes (the flux $P \vb{v}^{\psi_J} \ln f$ decays sufficiently rapidly at infinity, or suitable boundary conditions hold), only the last term survives, giving \eqref{eq:entropy_production}.

\section{Proof of \eqref{eq:N_B_evolution} and Verification of \eqref{eq:N_B_solution_intermediate}}
\label{app:proof_NB_evolution}
\label{app:derivation_NB_solution}

Integrating \eqref{eq:modified_continuity_born} over $\mathcal{C}$, the divergence term vanishes under sufficiently rapid decay of the current at infinity:
\begin{equation}
\frac{\dd N_B}{\dd t}
= \frac{2}{\hbar} \int_{\mathcal{C}} \dd \Sigma \,
\operatorname{Im}\bigl[ J e^{-\im S_J/\hbar} \bigr] R_J .
\label{eq:NB_integral_source_app}
\end{equation}
Substituting \eqref{eq:source_constraint_field},
\begin{equation}
\frac{\dd N_B}{\dd t}
= - \int_{\mathcal{C}} \dd \Sigma \,
\frac{|\psi_J|^2}{\tau_{\rm E}} \,
\frac{1 - f}{f}.
\label{eq:NB_intermediate_app}
\end{equation}
In the uniform case $f = f(t)$ and $\tau_{\rm E} = \tau$ factor out of the integral, giving \eqref{eq:N_B_evolution}.

For the solution, write $N_B(t) = N_B(\infty)[1 + (f_0 - 1)e^{-t/\tau}]^{-1}$. Then
\begin{equation}
\frac{\dd N_B}{\dd t}
= \frac{N_B(\infty)\,(f_0 - 1)\,e^{-t/\tau}}
{\tau\,\bigl[1 + (f_0 - 1)\,e^{-t/\tau}\bigr]^2}.
\end{equation}
On the other hand, using the uniform relaxation law
\(f(t)=1+(f_0-1)e^{-t/\tau}\) from
Section~\ref{sec:Uniform Relaxation and the Dynamics of Born Integral},
we have \(1-f=-(f_0-1)e^{-t/\tau}\) and
\(f=1+(f_0-1)e^{-t/\tau}\), so the right-hand side of
\eqref{eq:N_B_evolution} equals
\begin{equation}
-\frac{1}{\tau}\,\frac{1 - f}{f}\,N_B
= \frac{N_B(\infty)\,(f_0 - 1)\,e^{-t/\tau}}
{\tau\,\bigl[1 + (f_0 - 1)\,e^{-t/\tau}\bigr]^2},
\end{equation}
which agrees. The limit $t \to \infty$ gives $N_B(\infty)$ by definition, and at $t=0$ the solution gives $N_B(0) = N_B(\infty)/f_0$; normalization of $P = f_0|\psi_J|^2$ fixes $N_B(0) = 1/f_0$, so $N_B(\infty) = 1$. Since \eqref{eq:N_B_evolution} is a first-order linear ODE in $N_B$, the solution is unique.

\section{Non-Uniqueness of the Relaxation Source and Phase-Amplitude Decoupling}
\label{app:relaxation_nonuniqueness}

This appendix elaborates Remark~\ref{rem:relaxation_nonuniqueness}. The relaxation constraint \eqref{eq:source_constraint_field} reads
\begin{equation}
\operatorname{Im}\!\left[J(\vb{x},t) e^{-\im S_J(\vb{x},t)/\hbar}\right] = -\frac{\hbar R_J(\vb{x},t)}{2\tau_{\rm E}(\vb{x},t)} \frac{1-f(\vb{x},t)}{f(\vb{x},t)} =: \mathcal{C}(\vb{x},t).
\end{equation}
This is a single real equation for the two real components of $J$, and it constrains only the phase-rotated imaginary part. We introduce an arbitrary real scalar field $\lambda(\vb{x},t)$ such that
\begin{equation}
\operatorname{Re}\!\left[J(\vb{x},t) e^{-\im S_J(\vb{x},t)/\hbar}\right] = \lambda(\vb{x},t).
\label{eq:I_real_unconstrained}
\end{equation}
Then $J e^{-\im S_J/\hbar} = \lambda + \im\,\mathcal{C}$, and multiplying by $e^{\im S_J/\hbar}$ and using $\psi_J = R_J e^{\im S_J/\hbar}$,
\begin{equation}
J(\vb{x},t) = \frac{\lambda(\vb{x},t) + \im\,\mathcal{C}(\vb{x},t)}{R_J(\vb{x},t)} \psi_J(\vb{x},t).
\label{eq:I_general_relaxation_source}
\end{equation}
This is the most general source satisfying the relaxation constraint for the prescribed profile. The proportional ansatz $J \propto \im\psi_J$ is the special case $\lambda = 0$.

To see that $\lambda$ and $\mathcal{C}$ control independent channels, substitute \eqref{eq:I_general_relaxation_source} into \eqref{eq:modified_HJ}. Since $\operatorname{Re}[J e^{-\im S_J/\hbar}] = \lambda$,
\begin{equation}
\frac{\partial S_J}{\partial t} + \frac{(\nabla S_J)^2}{2m} + V + Q^{\rm B}_J = -\frac{\lambda(\vb{x},t)}{R_J(\vb{x},t)}.
\label{eq:I_lambda_in_HJ}
\end{equation}
Thus $\lambda$ enters the phase dynamics as an effective potential, while $\mathcal{C}$ does not appear. In the continuity equation \eqref{eq:modified_continuity_born} the source term is
\begin{equation}
\frac{2}{\hbar}\operatorname{Im}\!\left[J e^{-\im S_J/\hbar}\right] R_J = \frac{2}{\hbar} \mathcal{C}(\vb{x},t) R_J(\vb{x},t),
\label{eq:I_C_in_continuity}
\end{equation}
which depends only on $\mathcal{C}$. The amplitude dynamics, and hence the transport equation for $f$, is controlled exclusively by $\mathcal{C}$, and the phase dynamics, hence the velocity field $\vb{v}^{\psi_J} = \nabla S_J/m$, exclusively by $\lambda$. Any choice of $\lambda$ leaves the prescribed relaxation unchanged, and any choice of $\mathcal{C}$ leaves the trajectory dynamics unchanged at the level of the source terms.

The entropy drift for the general relaxation source follows from \eqref{eq:entropy_production},
\begin{equation}
\frac{\dd S_{\rm e}}{\dd t} = \frac{2 k_{\rm B}}{\hbar} \int_{\mathcal{C}} \dd \Sigma \, \frac{\mathcal{C}(\vb{x},t)}{R_J(\vb{x},t)} P(\vb{x},t),
\end{equation}
which does not contain $\lambda$.

\section{Alternative Classicalization Route and Entropy Drift}
\label{app:alternative_classicalization}

This appendix elaborates Remark~\ref{rem:classicalization_nonuniqueness}. We start from a classical action $S$ satisfying $\partial_t S+(\nabla S)^2/(2m)+V=0$, and construct $\psi=R e^{\im S/\hbar}$ with amplitude $R$. Substitution into the Schr\"odinger operator gives $\im\hbar\partial_t\psi=\hat{H}\psi+J$, where $J=[-Q^{\rm B}R+\im\mu]\psi/R$ and $\mu=\hbar(2R)^{-1}[\partial_t(R^2)+\nabla\cdot(R^2\nabla S/m)]$. The real part of the phase-rotated source, $\operatorname{Re}[J e^{-\im S/\hbar}]=-Q^{\rm B}R$, cancels the Bohm potential for any $R$. The imaginary part, $\operatorname{Im}[J e^{-\im S/\hbar}]=\mu$, measures the failure of $R$ to satisfy the continuity equation $\partial_t(R^2)+\nabla\cdot(R^2\nabla S/m)=0$. The reverse Madelung slice is the special case $\mu=0$, i.e.\ the case in which this continuity equation holds. Only then is $J=-Q^{\rm B}\psi$, the Born continuity equation unmodified, and the entropy drift zero. For $\mu\neq0$ the source lies outside the reverse Madelung slice, the Born density is not conserved, and the entropy drift is generally nonzero.

In the extended framework the physical requirement is only that the total quantum potential vanish. As in \eqref{eq:zero_total_constraint}, this imposes the single real constraint
\begin{equation}
\operatorname{Re}\left[ J e^{-\im S_J/\hbar} \right] = -Q^{\rm B}_J R_J.
\end{equation}
The imaginary part is unconstrained. We introduce an arbitrary real scalar field $\mu(\vb{x},t)$,
\begin{equation}
\operatorname{Im}\left[ J e^{-\im S_J/\hbar} \right] = \mu(\vb{x},t),
\label{eq:H_imag_unconstrained}
\end{equation}
so that the general classicalizing source is
\begin{equation}
J(\vb{x},t) = \left[ -Q^{\rm B}_J(\vb{x},t) R_J(\vb{x},t) + \im \mu(\vb{x},t) \right] \frac{\psi_J(\vb{x},t)}{R_J(\vb{x},t)}.
\label{eq:H_general_class_source}
\end{equation}
In contrast with the reverse Madelung construction above, $\mu$ is no longer determined by $R$ through the continuity equation; it is a free field. For any $\mu \neq 0$ the source lies outside the reverse Madelung slice, yet the trajectories are strictly classical.

For a concrete choice, a constant $\mu<0$ would give a source $\propto -\im\,e^{\im S_J/\hbar}$ that drives $R_J$ to zero in finite time and makes the transport equation for $f$ singular. We choose instead
\begin{equation}
\mu(\vb{x},t) = -\frac{\hbar\gamma(\vb{x},t)}{2}R_J(\vb{x},t),
\label{eq:H_mu_regular}
\end{equation}
with $\gamma(\vb{x},t) > 0$ a bounded real field. Then
\begin{equation}
J_{\rm reg}(\vb{x},t) = \left[-Q^{\rm B}_J(\vb{x},t) - \frac{\im\hbar\gamma(\vb{x},t)}{2}\right]\psi_J(\vb{x},t).
\label{eq:H_J_reg}
\end{equation}
This source combines the Bohm-potential cancellation with an imaginary potential $-\im\hbar\gamma/2$, so it coincides formally with a gain/loss construction; the new content is its combination with trajectory classicalization and the normalized accounting below. A direct check: $J_{\rm reg}e^{-\im S_J/\hbar} = (-Q^{\rm B}_J - \im\hbar\gamma/2)R_J$, so $\operatorname{Re}[J_{\rm reg}e^{-\im S_J/\hbar}] = -Q^{\rm B}_J R_J$, which gives $Q^{\rm total}_J = 0$ and classical trajectories, while $\operatorname{Im}[J_{\rm reg}e^{-\im S_J/\hbar}] = -\hbar\gamma R_J/2 \neq 0$.

The resulting dynamics is regular. Substituting \eqref{eq:H_J_reg} into the transport equation gives $D f/D t = \gamma f$, and the Born integral obeys $\dd N_B/\dd t = -\langle\gamma\rangle_\rho N_B$, where $\langle\gamma\rangle_\rho := \int_{\mathcal{C}} \gamma|\psi_J|^2 \dd\Sigma / N_B$. Defining the normalized ratio $\hat{f} := f N_B$,
\begin{equation}
\frac{D \hat{f}}{D t} = \hat{f}\left(\gamma(\vb{x},t) - \langle\gamma\rangle_\rho\right).
\label{eq:H_hat_f_evolution}
\end{equation}
This is nonzero exactly when $\gamma$ is nonuniform with respect to the Born density. If $\gamma$ is spatially constant, $\hat{f}$ is conserved: writing $\psi_J = e^{-\gamma t/2}\psi_0$, where $\psi_0$ solves the $\gamma = 0$ problem, the phase and velocity field are unchanged, so the trajectories and the distribution $P$ are those of the reverse Madelung slice, and only the norm changes, $N_B(t) = e^{-\gamma t}N_0$. The drift $\dd S_{\rm e}/\dd t = -k_{\rm B}\gamma$ is then a normalization effect and not a change of the normalized statistics. A genuinely distinct regime arises when $\gamma$ varies in space: regions lose Born density at different rates, $P$ no longer tracks the normalized Born density $|\psi_J|^2/N_B$, and $\hat{f}$ evolves according to \eqref{eq:H_hat_f_evolution}.

To isolate this effect we introduce the entropy relative to the normalized Born density, $\tilde{S} := -k_{\rm B}\int_{\mathcal{C}} P\ln\hat{f}\,\dd\Sigma = S_{\rm e} - k_{\rm B}\ln N_B$, and call its rate the \emph{norm-corrected entropy drift}:
\begin{equation}
\frac{\dd\tilde{S}}{\dd t} = k_{\rm B}\left(\langle\gamma\rangle_\rho - \langle\gamma\rangle_P\right),
\label{eq:H_normalized_entropy_exchange}
\end{equation}
where $\langle\gamma\rangle_P := \int_{\mathcal{C}} \gamma P\,\dd\Sigma$. This vanishes for constant $\gamma$ and whenever $P = |\psi_J|^2/N_B$; otherwise it is nonzero and can have either sign. Since $\tilde{S}$ is a proper relative entropy (non-positive, vanishing only at $\hat{f} = 1$), the sign of its drift depends on how $\gamma$ is distributed under $P$ versus the Born density, and is not fixed by $\gamma > 0$.

Finally, the entropy drift of the source $J_{\rm reg}$ follows from \eqref{eq:entropy_production},
\begin{equation}
\frac{\dd S_{\rm e}}{\dd t}\bigg|_{J_{\rm reg}} = \frac{2 k_{\rm B}}{\hbar} \int_{\mathcal{C}} \dd \Sigma \, \frac{\mu}{R_J} P = -k_{\rm B}\langle\gamma\rangle_P < 0 .
\label{eq:H_entropy_alt_explicit}
\end{equation}
It contains both the normalization part $-k_{\rm B}\langle\gamma\rangle_\rho$ and the norm-corrected part \eqref{eq:H_normalized_entropy_exchange}.

\section*{References}


\end{document}